\documentclass[aps,prx,reprint,longbibliography,superscriptaddress]{revtex4-2}

\usepackage[utf8]{inputenc}

\usepackage{xr}

\usepackage{graphicx}
\usepackage{amsmath, amssymb, braket}
\usepackage{tikz}
\usepackage{booktabs}

\usepackage{color, xcolor}

\usepackage[colorlinks=true,urlcolor=blue,linkcolor=red,citecolor=blue]{hyperref}

\definecolor{Blue}{HTML}{1f77b4}
\definecolor{Orange}{HTML}{ff7f0e}

\usepackage{array}
\newcolumntype{C}{>{$}c<{$}}

\newcommand{\Prefix}{Figures} %
\newcommand{\plot}[2][]{%
	{%
		\centering
		\includegraphics[#1]{\Prefix/#2}%
	}%
}

\newcommand{\dv}[2]{\frac{\partial{} #1}{\partial{} #2}}
\newcommand{\ddv}[2]{\frac{\partial{}^2 #1}{\partial{} #2^2}}

\newcommand{\rarrow}{\xrightarrow{r \gg{} \xi}}

\def\CL{\mathrm{A}}

\def\Op{\hat{O}}

\def\dO{\overline{\delta{} O^2}}

\def\En{\varepsilon}
\def\Ha{\hat{\mathcal{H}}}
\def\ha{\hat{h}}

\def\DD{\partial^2_{\En}}

\def\DE{\mathrm{DE}}
\def\CE{\mathrm{CE}}
\def\K{\mathrm{K}}

\newcommand{\ToDelete}[1]{{\color{gray} [DELETED]}}

\makeatletter

\begin{document}

\title{Thermalization Dynamics in Closed Quantum Many Body Systems:\\ a Precision Large Scale Exact Diagonalization Study}

\author{Ivo A. Maceira}
\affiliation{Institute of Physics, \'{E}cole Polytechnique F\'ed\'erale de Lausanne (EPFL), CH-1015 Lausanne, Switzerland}
\author{Andreas M. L\"auchli}
\affiliation{Laboratory for Theoretical and Computational Physics, Paul Scherrer Institute, CH-5232 Villigen-PSI, Switzerland}
\affiliation{Institute of Physics, \'{E}cole Polytechnique F\'ed\'erale de Lausanne (EPFL), CH-1015 Lausanne, Switzerland}

\date{\today}
\begin{abstract}
	Using a Krylov-subspace time evolution algorithm, we simulate the real-time dynamics of translation invariant non-integrable finite spin rings to quite long times with high accuracy. We systematically study the finite-size deviation between the resulting equilibrium state and the thermal state, and we highlight the importance of the energy variance on the deviations. We find that the deviations are well described by the eigenstate thermalization hypothesis, and that the von Neumann entropy correction scaling is the square of the local operator scaling. We reveal also an area law contribution to the relaxed von Neumann entropy, which we connect to the mutual information between the considered subsystem and its immediate environment. We also find that local observables relax towards equilibrium exponentially with a relaxation time scale that grows \textit{linearly} with system length and is somewhat independent of the local operator, but depends strongly on the energy of the initial state, with the fastest relaxation times found towards one end of the overall energy spectrum. To contrast this behaviour we also study domain wall initial states, which exhibit clear diffusive behaviour, with a Thouless time scaling quadratically with the systems size, leading to a rather precise estimate for the diffusion constant for states in the vicinity of the middle of the energy spectrum.
\end{abstract}

\maketitle

\tableofcontents

\section{Introduction}%
\label{sec:introduction}

It is generally believed that generic isolated quantum many body systems relax to a thermal state, in the sense that subsystems become indistinguishable from subsystems of a thermal/microcanonical state~\cite{Neumann:1929, Goldstein.Lebowitz.ea:2010, Deutsch:1991, Srednicki:1994, Gogolin.Eisert:2016, D’Alessio.Kafri.ea:2016}. The temperature of the thermal state is controlled by the expectation value of the energy in the initial state. The matter of thermalization has received considerable attention numerically~\cite{Prosen:1999, Kollath.Lauchli.ea:2007, Manmana.Wessel.ea:2007, Rigol.Dunjko.ea:2008, Rigol:2009, Rigol:2009*1, Bartsch.Gemmer:2009, Banuls.Cirac.ea:2011, Torres-Herrera.Santos:2013, Sorg.Vidmar.ea:2014, Beugeling.Moessner.ea:2014, Beugeling.Moessner.ea:2015, Luitz.Bar-Lev:2016, Chen.Zhou.ea:2018, Bleicker.Stolze.ea:2020, Richter.Dymarsky.ea:2020, Russomanno.Fava.ea:2021, Lezama.Torres-Herrera.ea:2021, Pappalardi.Fritzsch.ea:2023}, but we believe that the current capacities of exact diagonalization methods have not been fully exploited to address this question, and higher precision results can be achieved. To our knowledge, an in-depth and comprehensive analysis of thermalization dynamics on a non-integrable model is lacking in the current literature.

While so far most studies reported compatibility of the numerical simulations with the theoretical expectations, here we undertake a rather comprehensive and quantitative study of the convergence of late-time quench observables towards the expectations of thermalization. Our study relies on the fact that temporal fluctuations at late time are suppressed exponentially in the system size, therefore making a push for large system sizes very valuable.

In non-integrable models, the equilibrium state of typical quantum states undergoing relaxation after a quench is given by the diagonal ensemble, which can be obtained from a full diagonalization of the Hamiltonian~\cite{Prosen:1999, Prosen.Znidaric:2007, Kollath.Lauchli.ea:2007, Rigol.Dunjko.ea:2008, Rigol:2009, Rigol:2009*1, Bartsch.Gemmer:2009, Banuls.Hastings.ea:2009, Banuls.Cirac.ea:2011, Torres-Herrera.Santos:2013, Khatami.Pupillo.ea:2013, Sorg.Vidmar.ea:2014, Beugeling.Moessner.ea:2014, Beugeling.Moessner.ea:2015, Luitz.Bar-Lev:2016, Richter.Dymarsky.ea:2020, Banuls.Heller.ea:2020, Russomanno.Fava.ea:2021, Pappalardi.Fritzsch.ea:2023}. Diagonal ensemble studies thus provide complete accuracy, but the accessible system sizes may be too small to observe thermalization. While matrix product state simulations clearly outperform exact diagonalizations in system sizes~\cite{Kollath.Lauchli.ea:2007, Manmana.Wessel.ea:2007, Sorg.Vidmar.ea:2014, Chen.Zhou.ea:2018, Javier-Valencia-Tortora.Calabrese.ea:2020, Morningstar.Hauru.ea:2022} at short times after the quench, and also at very long times, when the system is believed to be described as a system slightly perturbed away from thermal equilibrium, here we instead investigate the dynamics of a closed quantum many body system over the entire dynamically relevant time scales for a finite size system.

In this paper we use a Krylov-subspace time evolution algorithm~\cite{Park.Light:1986} to simulate the real-time dynamics of finite systems~\cite{Prosen:1999, Lezama.Torres-Herrera.ea:2021} to quite long times with high accuracy and relatively large system sizes, considerably larger than those accessible with full (complete) diagonalization methods. Temporal averages on the equilibrated state are good estimates of the diagonal ensemble averages and we compare them to canonical ensemble thermal averages (obtained from exact diagonalization) at the corresponding energy density.

Taking a well-studied non-integrable spin chain as our Hamiltonian, we perform the time evolution on a large set of translationally invariant product states covering a wide range of energies and energy variances on chains of up to $L=34$ spins in one instance and $L=30$ for most initial states, allowing us to probe thermalization dynamics on a wide range of system parameters.

We find that the equilibrium state converges to the thermal state with finite-size deviations on local observables that are well described by the lowest order correction on $(v - \tilde{v})/L$ predicted by the eigenstate thermalization hypothesis (ETH)~\cite{Deutsch:1991, Srednicki:1994, Srednicki:1996, Srednicki:1999}, where $v$ and $\tilde{v}$ are the energy variance densities of the initial and thermal states respectively. While the lowest order correction has been known for some time and is simple to derive~\cite{D’Alessio.Kafri.ea:2016}, we find that it is seldom taken into account in the modern literature, and often the role of the energy variance is not considered in the interpretation of numerical results~\cite{Kim.Huse:2013, Zhang.Kim.ea:2015}. We estimate the constant prefactor on the lowest order correction from thermal expectations, letting us probe the agreement with the ETH at a higher quantitative level.

We also observe that the temporal fluctuations around equilibrium scale as the inverse of the density of states~\cite{Deutsch:1991, Srednicki:1994, Srednicki:1996, Srednicki:1999, Sorg.Vidmar.ea:2014, Nation.Porras:2019, Bleicker.Stolze.ea:2020, Knipschild.Gemmer:2020}, agreeing with the bound predicted by the ETH, and allowing the extraction of very precise estimates based on exact diagonalization at large system sizes.

We calculate the entanglement entropy of the equilibrium state on contiguous spin clusters and find that it is consistent with the corresponding thermal entropy on the same clusters. The initial state most similar to the infinite temperature thermal state ($Y_+$) achieves the maximum entropy, and this maximum agrees with the average entropy of random states uniformly drawn on the symmetry sector that the initial state belongs to.

The finite-size deviation between entropies is found to depend on the energy variance as the square of $(v - \tilde{v})/L$, while a $\mathcal{O}(1)/L$ term is also observed. This dependency can be derived from the ETH as shown in a recent publication~\cite{Huang:2025, Huang:2024}, without requiring further assumptions or extensions~\cite{Rigol.Srednicki:2012, Luitz.Bar-Lev:2016, Dymarsky.Lashkari.ea:2018, Kaneko.Iyoda.ea:2020}. Beyond the volume law contribution to the entanglement entropy, we observe and explain an area-law term in the entropy associated to the mutual information between the block and its immediate environment at finite temperatures.

We also study the relaxation dynamics of local observables, and the large $L$ results show that equilibrium values are reached after an exponential decay~\cite{Nation.Porras:2019, Dymarsky:2022} with a relaxation time $\tau$ that is proportional to system size~\cite{Dymarsky:2022}, while we are not able to observe a power-law decay consistent with hydrodynamics~\cite{Lux.Muller.ea:2014, Khemani.Vishwanath.ea:2018, Werman.Chatterjee.ea:2019}. While energy correlators show the most clear exponential dependence, the relaxation time scale $\tau$ is somewhat independent of the observable considered but depends significantly on the energy density. The entanglement entropy also relaxes exponentially at long times, although its analysis was not as thorough as the local observables.

The paper is organized as follows: In Sec.~\ref{sec:model} we discuss the spin chain Hamiltonian and the sets of initial states. In Sec.~\ref{sec:methods} we discuss our numerical method and define relevant quantities and observables. In Sec.~\ref{sec:thermalization_of_local_observables} we analyze the scaling of the deviations between diagonal and thermal ensembles, we show the exponential decay of the temporal fluctuations with $L$, and we demonstrate the agreement of these quantities with the ETH. In Sec.~\ref{sec:thermalization_of_entanglement_entropy} we focus on the entanglement entropy and repeat the analysis on the deviations, and we observe the appearance of the area-law term in the entropy. In Sec.~\ref{sec:exponential_relaxation} we exemplify the exponential decay of observables over time and we study the dependence of the decay time scale $\tau$ on the observable, energy, energy variance, and system size.

In the main text we generally contrast two initial states, one corresponding to infinite temperature and another to finite temperature. The main text results are a representative subset of our full analysis which covers a wide range of system parameters, states, and observables. We present complementary results in appendices and supplemental material (SM).

\section{Model and initial states}%
\label{sec:model}

We consider isolated $S=1/2$ spin chains with a translationally invariant and non-integrable Hamiltonian:
\begin{align}
	\Ha   & = \sum_{j} \Ha_j,                                                                                                          \\
	\Ha_j & \equiv \sigma^z_j\sigma^z_{j+1} + \frac{h_x}{2}(\sigma^x_j + \sigma^x_{j+1}) + \frac{h_z}{2}(\sigma^z_j + \sigma^z_{j+1}),
	\label{eq:Hamiltonian}
\end{align}
where we defined local energy operators centered on the bonds. For generic field parameters, the Hamiltonian~\eqref{eq:Hamiltonian} is expected to have no additional symmetries beyond the spatial translation and reflection symmetries and time reversal symmetry.

This Hamiltonian is also called the mixed field Ising model, and beyond its significance as a prominent one-dimensional system to study non-equilibrium physics theoretically and numerically, it can also be implemented or is realized in natural~\cite{Coldea.Kiefer:2010} and synthetic~\cite{Browaeys.Lahaye:2020, Chen.Pan:2021} experimental systems.

This Hamiltonian was already considered in the context of thermalization, and we will work with the following parameter set for which the Hamiltonian~\eqref{eq:Hamiltonian} has been established (numerically) to be non-integrable~\cite{Banuls.Cirac.ea:2011, Chen.Zhou.ea:2018, Bonfim.Boechat.ea:2019, Banuls.Huse.ea:2020, Cakan.Cirac.ea:2021}:
\begin{equation}
	h_x=-1.05, \ h_z=0.5\ .
	\label{eq:BCH}
\end{equation}

\begin{figure}
	\centering
	\plot{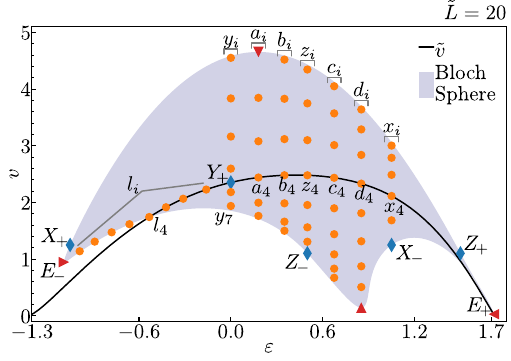}
	\caption{The Bloch sphere mapped to the $(\En, v)$ plane with the Hamiltonian~(\ref{eq:Hamiltonian}). The black line is the thermal energy variance density $\tilde{v}$. The initial states we time evolve are marked on the surface. The quantization axes states are marked with blue diamonds, and the extremes of energy and variance with red triangles. We categorize most positive energy states in equal energy series like the $z_i$ series which starts with $z_1$ and ends with the $Z_-$ state.}%
	\label{fig:v_epsilon}
\end{figure}

Following Ref.~\cite{Banuls.Cirac.ea:2011}, we will consider translationally invariant and reflection symmetric pure product states as initial states:
\begin{equation}
	|\psi_\mathrm{initial}\rangle = \prod_i |(\theta,\phi)\rangle_i
\end{equation}
Each initial state is then parametrized by the spherical angles $(\theta, \phi)$ that indicate where all spins point to initially. Each state maps to a point in the $(\En, v)$ plane, where $\En$ is the energy density and $v$ the energy variance density of the state with respect to the chosen Hamiltonian and its parameters. In Fig.~\ref{fig:v_epsilon} we show the image of the map from the surface of the Bloch sphere to the $(\En, v)$ plane and we mark the initial states we consider. See Appendix~\ref{sec:bloch_states} for additional details and exact definitions of the states.

Except for the quantization axes states $\{X_{\pm}, Y_{\pm}, Z_\pm\}$ and the extrema of energy $\{E_{\pm}\}$ within the manifold of initial states, all other states are organized in equal energy series (except $l_i$) and ordered by decreasing variance. Some states were purposefully defined such that their energy variance density $v$ is equal to the thermal energy variance density $\tilde{v}$ (black line) up to an error of $10^{-4}$: These are the fourth states in the positive energy series and $l_1$ to $l_4$ in the $l_i$ series. For $Y_{\pm}$ we have $v = \tilde{v}$ exactly.

The highest energy Bloch state $E_+$ is extremely close to the highest energy eigenstate which is a paramagnetic state lying close to $Z_+$~\cite{Bonfim.Boechat.ea:2019}. On the other hand the state $E_-$ is further from the ground state at $\En \approx -1.3$ which is a paramagnetic state with a significant overlap with an antiferro N\'eel state along the $z$-direction as the model parameters lie close to the antiferro-paramagnetic Ising CFT phase transition in the mixed field antiferromagnetic Ising chain ground state phase diagram.

\section{Methods and Observables}%
\label{sec:methods}

We employ a Krylov-subspace time integration method to approximate the propagator $\exp(-i\Ha t)$~\cite{Park.Light:1986, Hochbruck.Lubich:1997}. This method is basically numerically exact for the finite systems considered, i.e.~there is no truncation of the Hilbert space, and the time discretization error can be made arbitrarily small irrespective of the time step size. This allows us to accurately simulate the real-time dynamics of finite systems to quite long times. The limitations of this approach are the exponential growth of resource requirements with system size and the linear cost in final times reached. We perform the time evolution in the zero momentum and reflection symmetric sector that our initial states belong to. We time evolved the $Y_+$ state at system sizes up to $L=34$ spins, $Z_-$ up to $32$ spins, and other states up to $30$ or $28$ spins. Simulations were carried out to final times of $t_f=100$ with time steps of $0.5$ for $L \geq 32$ and $0.1$ otherwise, and measurements of observables or the entropy are carried out at each time step. Note that $t_f$ is much smaller than the Heisenberg time of the system (c.f. Appendix~\ref{sec:heisenberg_time}).

The expectation $\braket{\Op}_{\DE}$ of an observable on the diagonal ensemble is defined as the time independent component of the matrix element
\begin{align}
	\braket{\Psi|\Op(t)|\Psi} & = \braket{\Op}_{\DE} + \delta O (t),                 \\
	\braket{\Op}_{\DE}        & \equiv \sum_{j} |c_j|^2 \bra{\En_j} \Op \ket{\En_j},
\end{align}
where $\ket{\Psi} = \sum_{j} c_j \ket{\En_j}$ and $\ket{\En_j}$ are eigenstates of energy density $\En$. We estimate the diagonal ensemble expectation by time-averaging the time-evolved matrix element, ignoring the initial period of relaxation:
\begin{equation}
	\braket{\Op}_{\DE} \approx \bar{O} \equiv \frac{1}{t_m} \int_{t_f-t_m}^{t_f} \braket{\Psi|\Op(t)|\Psi} dt,
	\label{eq:diagonal_ensemble_time_average}
\end{equation}
where we take $t_m = t_f/4 = 25$, i.e.~the last quarter of the simulated time window.

We denote time averages with an overline and thermal expectations with a tilde. We will compare the diagonal ensemble estimate $\bar{O}$ of a state of average energy density $\En$ to the thermal average $\tilde{O}$ of the observable $\Op$ at the inverse temperature $\beta(\En)$.

Another quantity we study is the average of the fluctuations around the relaxation value $\bar{O}$, which we may calculate as
\begin{equation}
	\dO \equiv \frac{1}{t_m}\int_{t_f-t_m}^{t_f} dt\braket{\Op(t)}^2 - \bar{O}^2,
	\label{eq:delta_Ot}
\end{equation}
where we will take an average over the same time interval as in Eq.~(\ref{eq:diagonal_ensemble_time_average}). The ETH gives an upper bound on the fluctuations~\cite{Srednicki:1999, D’Alessio.Kafri.ea:2016}:
\begin{equation}
	\overline{\delta O^2} \leq M 2^{-\tilde{S}(\En)},
	\label{eq:delta_Ot_ETH}
\end{equation}
for some constant $M$. In Sec.~\ref{sub:fluctuations_around_equilibrium} we will see that this bound is close to being saturated.

The canonical thermal averages are obtained from the complete spectrum and eigenstates of the Hamiltonian~(\ref{eq:Hamiltonian}), which was obtained for system sizes up to $\tilde{L} = 20$. $\tilde{L}$ will refer to the system size used to extract thermal averages, but this parameter is less relevant than $L$ as thermal averages converge faster in system size for the considered energy densities $\epsilon$ (See for example Fig.~\ref{fig:sum_delta_C_r} in Appendix~\ref{sec:conservation_of_energy_variance}).

Apart from a selection of Pauli operators, we additionally focus on the energy correlators as our observables:
\begin{equation}
	\hat{C}_r(t) \equiv (\Ha_0(t)-\En)(\Ha_r(t)-\En).
\end{equation}
They feature relatively small temporal fluctuations, are the most sensitive to the energy variance, and show the best agreement with the ETH scaling predictions, and in particular $\hat{C}_2$ showed the clearest exponential dynamics of all, so we focused on this operator more than others.

We also calculate the entanglement entropy
\begin{equation}
	S_l \equiv - \mathrm{Tr}( \rho_l \log_2 \rho_l ),
\end{equation}
of a contiguous spin cluster of length $l=1,2,3$, where $\rho_l$ is the reduced density matrix, up to $L=34$ spins and $l\leq L/2$ for $L=24$. In Sec.~\ref{sec:thermalization_of_entanglement_entropy} we compare $S_l$ to the rescaled thermal entropy $l\tilde{S}/\tilde{L}$, where $\tilde{S} \equiv - \mathrm{Tr}( \tilde{\rho} \log_2 \tilde{\rho} )$, and to the von Neumann entropy $\tilde{S}_l \equiv - \mathrm{Tr}( \tilde{\rho}_l \log_2 \tilde{\rho}_l )$, where $\tilde{\rho}_l$ is the reduced thermal density matrix.

\section{Numerical Results: Thermalization of local observables}%
\label{sec:thermalization_of_local_observables}

\begin{figure}
	\centering
	\plot{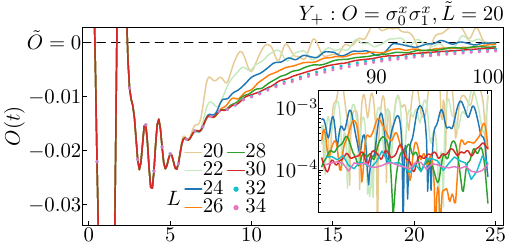}
	\put(-207, 108){(a)}

	\plot{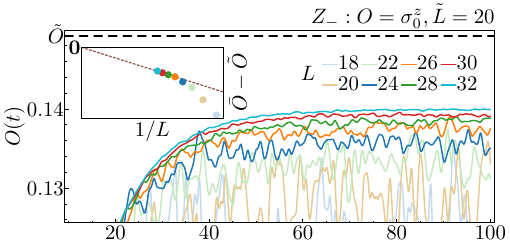}
	\put(-207, 108){(b)}

	\plot{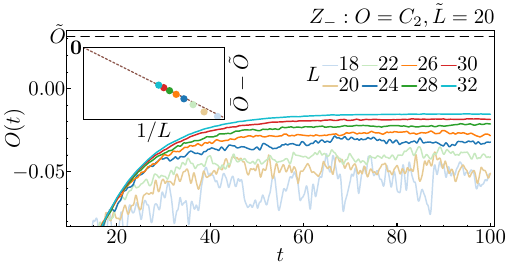}
	\put(-207, 119){(c)}

	\caption{Expectation of local operators over time on two mid-spectrum states. (a)~Effectively all local observables decay towards the infinite temperature expectation for the $Y_+$ state since $v = \tilde{v}$ for this state. In the inset we plot $|O(t)|$ in lin-log scale at the final simulation times. The temporal fluctuations decrease exponentially with system size, as demonstrated in Sec.~\ref{sub:fluctuations_around_equilibrium}. (b, c)~The expectation at equilibrium of $Z_-$ is observed to be offset from the thermal expectation since $v \neq \tilde{v}$, but it drifts towards this value as $\sim 1/L$, as shown in the inset. The dotted brown lines are visual guides crossing the origin (bold $0$).}%
	\label{fig:O_t}
\end{figure}

As discussed in the introduction, one can study the dynamics of thermalization in two distinct limits: One is to take the infinite system limit first $L\rightarrow\infty$, and then to study the dynamics of local observables as a function of time $t$. This is usually the setting in which MPS time evolution methods work. The second limit is to keep $L$ fixed first, then to simulate the dynamics upon convergence in time $t$ (formally $t\rightarrow \infty$), and then to study the finite size effects as the infinite system limit is approached. We take the second approach, and we start by describing the time evolution of local observables over time on the range of accessible system sizes.

\subsection{Observation of Relaxation}%

Given our observations, we describe the time evolution of a local observable in two phases: In a first phase, all the initially coherent phases in the eigenstate decomposition of the product state dephase~\cite{Kiendl.Marquardt:2017, Oliveira.Charalambous.ea:2018} and the observable evolves abruptly with large fluctuations (Fig.~\ref{fig:O_t}a). At such early times the system has not yet recognized it is finite and different system size curves overlap up to a time related to a Lieb-Robinson bound or diffusion constant.

We dedicate the analysis in this paper to what we consider as the second phase of time evolution: In this phase, the observable relaxes in a system-size dependent fashion and it decays towards an equilibrium value around which it fluctuates. The size of the fluctuations seems to be constant in time throughout the remainder of the time evolution. We expect the equilibrium value to be approximately the expectation on the diagonal ensemble, which might not be equal to the thermal expectation but approaches it as $1/L$ goes to zero, as is apparent in the insets of Fig.~\ref{fig:O_t}b,c.

The decay is dependent on $L$ and smaller systems reach equilibrium faster, but
with significantly larger temporal fluctuations. We demonstrate in Sec.~\ref{sec:exponential_relaxation} that the decay in the second phase is exponential and we explore the scaling of the corresponding time scale.

\begin{figure}
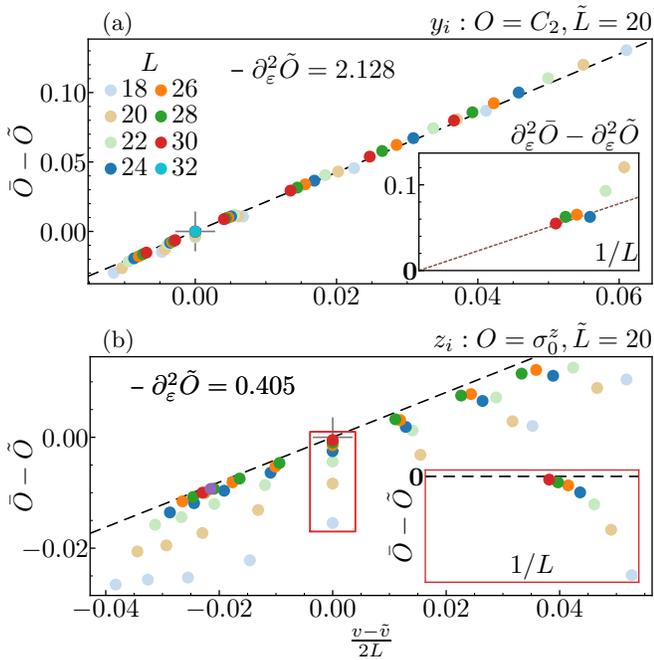

	\centering
	\plot{O-Lv,L,psi=y_i,O=C2.pdf}
	\put(-210, 108){(a)}

	\plot{O-Lv,L,psi=z_i,O=Sz0.pdf}
	\put(-210, 119){(b)}
	\caption{Difference between equilibrium average $\bar{O}$ and the thermal expectation of local observables for two series of equal energy states, as a function of the scaling parameter in Eq.~(\ref{eq:ETH_scaling}). The black dashed line has slope equal to $\DD \tilde{O}$ and goes through the origin (gray plus). The agreement is quite clear for $O=C_2$. (a) We take the $y_i$ series of states ($y_1$ to $y_7$ in Fig.~\ref{fig:v_epsilon}). The dotted brown line is a visual guide going through the origin $\mathbf{0}$. (b) We take the finite temperature $z_i$ series of states. Note the larger range of the vertical axis on the main panel of (a) compared to (b). The inset shows the deviation for the $y_4$ state (red box, $v \approx \tilde{v}$) as a function of $1/L$.}%
	\label{fig:ETH_scaling}
\end{figure}

\subsection{Structure of finite size and variance corrections}%

Comparing the expectation of the energy correlator $\hat{C}_2$ with the local spin operator $\hat{\sigma}^z$ (Fig.~\ref{fig:O_t}b,c) we see how $\hat{C}_2$ has larger deviations from the thermal value and a better agreement with a deviation scaling of $1/L$ (inset plots). The $1/L$ scaling is expected from the ETH: As discussed in Appendix~\ref{sec:ETH_prediction} in detail, assuming the ETH ansatz for the matrix elements of local operators, and given that both the thermal and diagonal ensembles asymptotically converge to a normal distribution~\cite{Hartmann..ea:2004}, we can calculate the finite size difference between thermal and diagonal expectations as a series in $1/L$, giving in lowest order
\begin{equation}
	\bar{O} - \tilde{O} \approx \frac{v-\tilde{v}}{2L} \DD O(\En) + \mathcal{O}\left( \frac{v^2-\tilde{v}^2}{L^2} \right),
	\label{eq:ETH_scaling}
\end{equation}
where $\DD O(\En)$ is the second derivative of $O(\En)$ at the mean energy density $\En$ and $O(\En)$ is the expectation of the local operator $\Op$ in the thermodynamic limit which according to the ETH is a smooth function only dependent on the mean energy. While Eq.~\ref{eq:ETH_scaling} is a known result~\cite{Rigol.Dunjko.ea:2008}, it has not been properly explored numerically, and it turns out its verification can go beyond an asymptotic scaling check since the prefactor $\DD O(\En)$ can be accurately estimated from thermal averages by expanding $\DD \tilde{O}(\En)$, the second derivative of the thermal expectation $\tilde{O}(\En)$, as we show in Appendix~\ref{sec:derivatives_of_o_en}.

We verify the scaling~(\ref{eq:ETH_scaling}) in Fig.~\ref{fig:ETH_scaling} by taking the equilibrium average $\bar{O}$ from series of initial states of equal energy but different energy variance. The agreement of our data with Eq.~(\ref{eq:ETH_scaling}) is clear for the energy correlator $\hat{C}_2$ (Fig.~\ref{fig:ETH_scaling}), possibly because of the large prefactor $\DD O(\En)$ for this operator.

Even for a finite temperature series of states like $z_i$, $\hat{C}_2$ still shows excellent agreement (See Fig.~\ref{fig:Obar_Lv_C2}). In contrast, other finite size effects commonly appear for spin operators, as seen for $\hat{\sigma}^z$ in Fig.~\ref{fig:ETH_scaling}b, although the finite size drift towards the predicted line seems to indicate an asymptotic agreement. Note that the discrepancy observed cannot be explained by considering higher order terms in Eq.~(\ref{eq:ETH_scaling}) as those still predict an agreement when $v = \tilde{v}$ which is not observed. It's possible that higher order moments of the energy are still significant for this operator at the system sizes considered. The discrepancy could also be due to relevant (but exponentially vanishing) off-diagonal matrix elements of the operator, which would be in accordance with the fast approach to zero in the inset of Fig.~\ref{fig:ETH_scaling}b.

We linearly fit the data shown in Fig.~\ref{fig:ETH_scaling}a for each system size to estimate a size dependent prefactor $\DD \bar{O}$, which we compare to the thermal estimate in the inset plot. The data is compatible with an agreement between these quantities at $L \rightarrow \infty$. The analysis of the prefactors $\DD \bar{C}_r$ is extended in Appendix~\ref{sec:conservation_of_energy_variance}.

\subsection{Fluctuations around equilibrium}%
\label{sub:fluctuations_around_equilibrium}

\begin{figure}
	\centering
	\plot{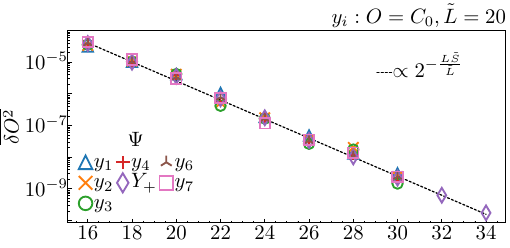}
	\put(-205, 108){(a)}

	\plot{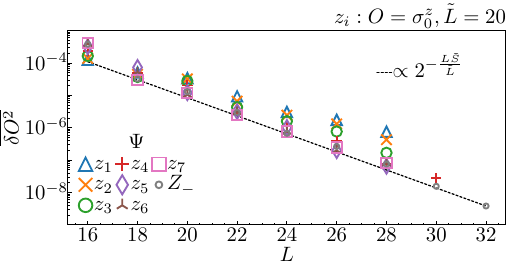}
	\put(-205, 119){(b)}
	\caption{Average size of the temporal fluctuations (Eq.~(\ref{eq:delta_Ot})). The dotted black lines follow the right-hand side prediction of Eq.~(\ref{eq:delta_Ot_ETH}), where the prefactor $M$ was arbitrarily chosen for the line to match the largest $L$ data points. }%
	\label{fig:delta_O}
\end{figure}

We calculate the temporal fluctuations $\overline{\delta O^2}$ of local observables as defined in Eq.~(\ref{eq:delta_Ot}) and plot them as a function of $L$ in Fig.~\ref{fig:delta_O} for the $y_i$ and $z_i$ series of states. The fluctuations decay exponentially in system size with a rate of decay that is approximately equal to the thermal entropy per site $\tilde{S}/\tilde{L}$ of the system, meaning that the inequality~(\ref{eq:delta_Ot_ETH}) is close to an equality. Note that it is this exponential suppression of the temporal fluctuations with system size which renders exact diagonalization of large systems a powerful technique to analyze non-equilibrium dynamics at late times.

The agreement is very good for the infinite temperature series $y_i$, while there seems to be some state dependence of the fluctuations at finite temperature, which is clearer in the local spin expectations, as seen in Fig.~\ref{fig:delta_O}b.

The prefactor $M$ in Eq.~(\ref{eq:delta_Ot_ETH}) is possibly dependent on the energy variance of the state, as the fluctuations are overall larger for the higher variance states.

\section{Numerical Results: Thermalization of entanglement entropy}%
\label{sec:thermalization_of_entanglement_entropy}

In this section we study the behaviour of the entanglement entropy of short subsystems of contiguous spins for long times. It is expected by now that the entanglement entropy converges to the thermal entropy of the same subsystem at late times, but a quantitative study of the finite size and finite time convergence properties has not be conducted yet. Furthermore much of the discussion has focused on the volume law part of the entanglement entropy, while we also reveal and discuss an area-law subleading part which is closely connected to the mutual information of the block with its immediate environment. Since the mutual information is zero at infinite temperature for the considered spin chain, this term only starts to appear for initial states which correspond to finite temperature.

\begin{figure}
	\centering

	\plot{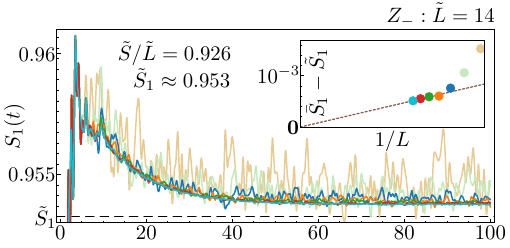}
	\put(-213, 108){(a)}

	\plot{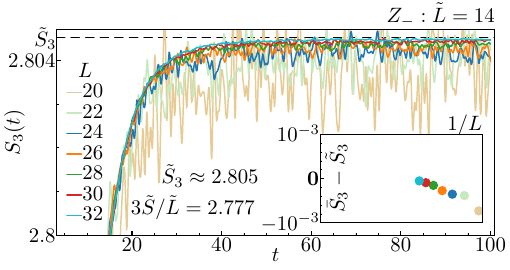}
	\put(-213, 119){(b)}

	\caption{The entanglement entropy of the $Z_-$ state on $1$ and $3$ site spin clusters. The inset plots compare the equilibrium average $\bar{S}_l$ with the thermal entropy $\tilde{S}_l$ of reduced density matrices on $l$ sites. The brown visual aid crosses the origin $\mathbf{0}$. We discuss the discrepancy between the values of $\tilde{S}_l$ and the scaled entropy $l\tilde{S}/\tilde{L}$ in Sec.~\ref{sub:mutual_information_and_area_law_term}.}%
	\label{fig:S_t_Zm}
\end{figure}

\subsection{Temporal behavior and approach to plateau value}%

The entanglement entropy $S_l$ of the product states grows from zero to an equilibrium plateau at late times~\cite{Kim.Huse:2013, Zhang.Kim.ea:2015} through what we also describe as two temporal phases: The initial growth of entropy is independent of the cluster size up to values close to the plateau and is expected to be linear in time~\cite{Lauchli.Kollath:2008,Kim.Huse:2013, Fagotti.Collura:2015, Znidaric:2020, Chang.Chen.ea:2019, De-Palma.Hackl:2022, Coppola.Tirrito.ea:2022}. Then the time evolution of $S_l$ is subject to an exponential relaxation towards an equilibrium plateau value $\bar{S}_l$ which is offset from the thermal value, but seems to converge in system size to the thermal entropy $\tilde{S}_l$ of a cluster of $l$ spins, as shown in Fig.~\ref{fig:S_t_Zm} for the $Z_-$ state.

The entanglement entropy of $Y_+$ (the maximum entropy state) at equilibrium is exponentially close to the maximum $l$ (in units of $\log_2$), and agrees with the expected entropy of random states drawn from the symmetry sector of $Y_+$. We discuss this in detail in Appendix~\ref{sec:entropy_of_y}.

Due to large fluctuations of the entropy in the initial phase, we could not obtain accurate estimates of a linear growth. The exponential regime that follows is shown for the $Z_-$ state in Fig.~\ref{fig:S_t_Zm} and in Fig.~\ref{fig:logS_t} on a log scale. The exponential decay is explored further in Appendix~\ref{sec:relaxation_of_entanglement_entropy}.

\subsection{Finite size and variance of initial states effects}%

\begin{figure}
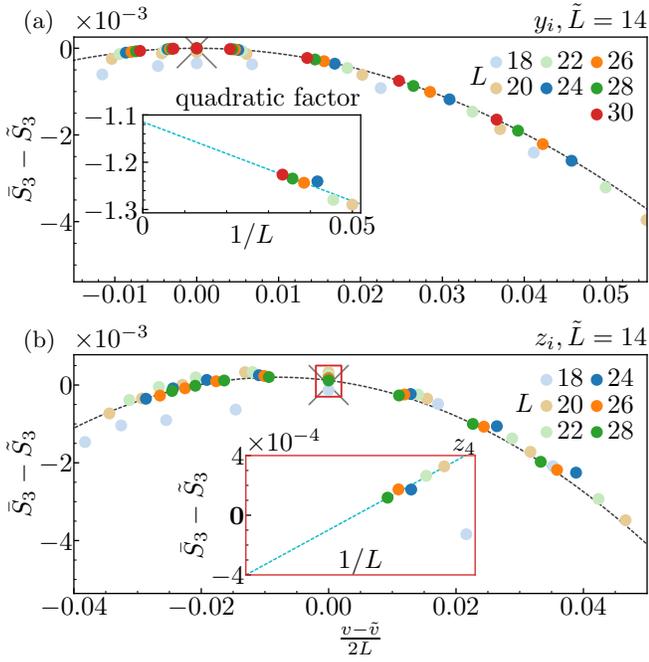

	\centering

	\plot{O-Lv,L,psi=y_i,O=S0,1,2.pdf}
	\put(-238, 108){(a)}

	\plot{O-Lv,L,psi=z_i,O=S0,1,2.pdf}
	\put(-238, 118){(b)}

	\caption{The deviation between the temporal average of the entanglement entropy and the entropy of the reduced thermal density matrix, as a function of the scaling parameter from Eq.~(\ref{eq:ETH_scaling}). The blue dotted lines in the inset plots are results of linear fits of the data. (a) Infinite temperature case: A quadratic dependence is apparent. For each $L$ we fit the data to a quadratic polynomial and plot in the inset the quadratic factor. (b) A quadratic dependence is still visible but other finite size effects might be relevant. The finite temperature maximum of entanglement entropy is not observed at $v = \tilde{v}$.}%
	\label{fig:S_Lv}
\end{figure}

The inset in Fig.~\ref{fig:S_t_Zm}a suggests the entanglement entropy converges towards the thermal entropy of the same subsystem as $1/L$. In Fig.~\ref{fig:S_t_Zm}b, it is unclear whether convergence is achieved below an error of $10^{-3}$ due to irregular finite size effects. Note also that the finite size effects on the thermal entropy $\tilde{S}_l$ could be relevant as we have taken a smaller $\tilde{L}$ due to the increased cost of calculating the reduced density matrices for all eigenstates. The same can be said about the deviation for $z_4$ as plotted in the inset of Fig.~\ref{fig:S_Lv}b, where the error in the $L\rightarrow \infty$ limit is again of order $10^{-3}$.

At the same time, the finite size scaling of the entropy deviation is not as simple as for local operators: We observe both $1/L$ and $1/L^2$ relevant terms at the scales studied, as shown in Fig.~\ref{fig:S_Lv}, with the second order term being proportional to $(v-\tilde{v})^2/L^2$, the square of the ETH scaling parameter of local operators (Fig.~\ref{fig:ETH_scaling}). Note that this does not coincide with the second order term of Eq.~(\ref{eq:ETH_scaling}). A quadratic dependence was expected as only $Y_+$ can achieve the maximum entropy, but the ETH formalism in Appendix~\ref{sec:ETH_prediction} does not explain this exact dependence. However, soon after our posting on arXiv of this numerical observation, the quadratic dependency was derived for our context assuming only the ETH~\cite{Huang:2025}, by developing previous calculations on the approximation of reduced density matrices~\cite{Huang:2024}. In~\cite{Huang:2025}, the quadratic prefactor was estimated as $\sim 1.18$, which is not far from the quadratic factors resulting from fitting (inset of Fig.~\ref{fig:S_Lv}a).

\subsection{Mutual information and area-law term}%
\label{sub:mutual_information_and_area_law_term}

\begin{figure}
	\centering
	\plot{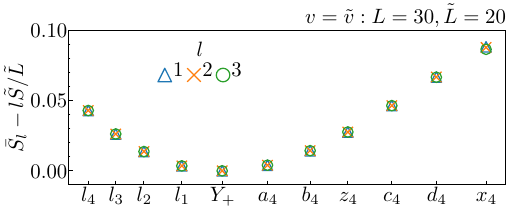}
	\put(-210, 90){(a)}

	\plot{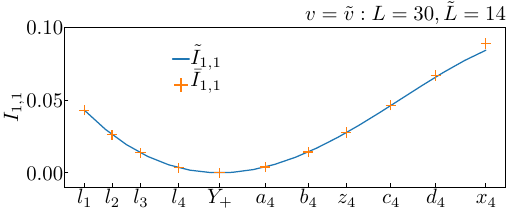}
	\put(-210, 90){(b)}

	\caption{(a) Entanglement entropy relative to the thermal entropy of many thermal variance states, as a function of their energy density. The independence on cluster size $l$ indicates the presence of an area-law term ($G$ in Eq.~(\ref{eq:area_law})). (b) The area-law term appears on the mutual information associated to the entropies $\bar{S}_l$ and $\tilde{S}_l$, which we see are approximately equal: The difference $\bar{I}_{1,1} - \tilde{I}_{1,1}$ is smaller than $10^{-3}$ except for $d_4$ and $x_4$. The area-law term $G$ depends approximately quadratically in $\En$ around $\En = 0$. The full data is provided in Table~\ref{tab:I_l_l}.}%
	\label{fig:I_E}
\end{figure}

As is shown by the indicated values in Figs.~\ref{fig:S_t_Zm}(a) and (b) we observe an offset between the late time entanglement entropy of a subsystem and the thermal entropy density times the volume of the subsystem. This excess of entanglement entropy with respect to the thermal entropy per spin is almost independent of the cluster size $l$ for small $l$ in the considered initial energy window (Fig.~\ref{fig:I_E}a), so we may write
\begin{equation}
	\bar{S}_l \approx \tilde{S}_l \approx l\tilde{S}/\tilde{L} + G,
	\label{eq:area_law}
\end{equation}
for $l \lesssim L/2$, where $G$ is an area-law term which is representative of increasing correlations in both the canonical ensemble at finite temperature~\cite{Wolf.Verstraete.ea:2008} and our late-time relaxed pure state.

Ref.~\cite{Deutsch:2010} argued that the entanglement entropy per spin of a mid-spectrum state should agree approximately with the thermal entropy per spin. However, the thermal entropy density alone fails to account for the emerging correlations between spin clusters at finite temperature~\cite{Wolf.Verstraete.ea:2008}, leading to a subleading area law term. The correlations contribute to the mutual information $I_{\mathrm{A}}$ between a contiguous cluster A and its complement~\cite{Wolf.Verstraete.ea:2008, Banuls.Yao.ea:2017, Alba.Calabrese:2019}, defined as
\begin{equation}
	I_{\CL} \equiv S_{\CL} + S_{\bar{\CL}} - S_{\CL \cup \bar{\CL}},\quad S_{\CL \cup \bar{\CL}} = 0.
\end{equation}
Furthermore, we can isolate the contribution of correlations to the mutual information by limiting the complement cluster to a surrounding neighborhood of size comparable or only somewhat larger than the correlation length (Fig.~\ref{fig:cluster_diagram}), thus avoiding contributions related to the pure state nature of the quenched state (Fig.~\ref{fig:I_l}) or possible terms scaling as the square root of the length~\cite{Murthy.Srednicki:2019} (which appear for clusters with $l \sim L/2$).
\begin{figure}
	\centering
	\begin{tikzpicture}
		\def\bh{0.4}
		\draw[Orange, very thick] (4.45, -\bh) rectangle (5.95, \bh);
		\node[below] at (5.2,-\bh){$B_2$};
		\draw[Orange, very thick] (1.25, -\bh) rectangle (2.75, \bh);
		\node[below] at (2,-\bh){$B_1$};
		\draw[Blue, very thick] (2.8, -\bh) rectangle (4.4, \bh);
		\node[below] at (3.6,-\bh){$A$};
		\foreach \x in {0, 0.8, ..., 7.201}
			{\fill[black] (\x, 0) circle (0.13);}
	\end{tikzpicture}%
	\label{fig:cluster_diagram}
	\caption{In our setting the correlation length is small so a spin cluster $A$ is mainly correlated with a small neighborhood cluster $B = B_1 \cup B_2$. The correlations are then quantified by the mutual information $I_{A,B} \approx 2I_{A,B_1}$ between $A$ and $B$. In the case shown $\ell=2$.}%
\end{figure}
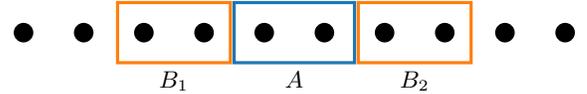
\begin{figure}
	\centering
	\plot{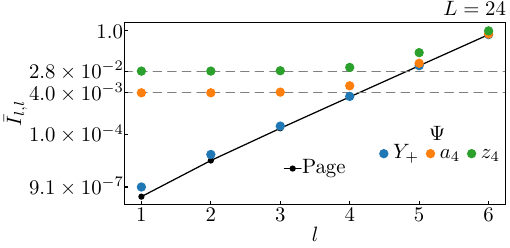}
	\caption{The time-averaged mutual information $\bar{I}_{l,l}$ between two equal-sized clusters of three thermal variance states as a function of $l$, compared with the expected mutual information of a random state (Page, Eq.~(\ref{eq:page_MI})). The dashed lines cross the $\bar{I}_{1,1}$ values for each state. The $Y_+$ mutual information does not exactly agree with the Page expectation because the state belongs to a symmetry sector of the Hamiltonian (see Appendix~\ref{sec:entropy_of_y}).}%
	\label{fig:I_l}
\end{figure}
As a further simplification, we calculated the mutual information between two contiguous clusters of sizes $l$ and $r$,
\begin{equation}
	I_{l,r} \equiv S_l + S_{r} - S_{l+r},
\end{equation}
which in this sense will quantify approximately half the correlations between a cluster of size $l$ and all spins at a distance $\leq r$ from it. This quantity becomes equal to the area-law term at equilibrium,
\begin{equation}
	\bar{I}_{l,r} \approx \tilde{I}_{l,r} \approx A,
\end{equation}
for $l+r \ll L/2$ (Fig.~\ref{fig:I_E}b). The quantum nature of the state becomes increasingly relevant at larger $l$. This is clear for an infinite temperature ($A=0$) random state: The expected entanglement entropy of such a state uniformly drawn according to the Haar measure, as first estimated by Page~\cite{Page:1993}, is known exactly~\cite{Foong.Kanno:1994, Bianchi.Dona:2019} and differs from the thermal entropy maximum $l$:
\begin{equation}
	P_l \sim l - \frac{2^{2l-L}}{2\log 2} + \mathcal{O}(2^{l-L})
	\label{eq:page_entropy}
\end{equation}
for $l \leq L/2$. The mutual information $I_{l, l}$ associated to the Page entropy (Eq.~(\ref{eq:page_entropy})) is then
\begin{equation}
	I_{l,l} \sim \frac{2^{4l-L}}{2\log 2} + \mathcal{O}(2^{2l-L})
	\label{eq:page_MI}
\end{equation}
for $l \leq L/4$. Thus a random state will have a macroscopic component of $\bar{I}_{l,l}$ at larger $l$ while the thermal counterpart is zero.

We plot Eq.~(\ref{eq:page_MI}) in Fig.~\ref{fig:I_l} together with the mutual information of three states. The mutual information of $Y_+$ grows exponentially with $l$ up to $l = L/4$ and is very close to the Page expectation (Eq.~(\ref{eq:page_MI})). At finite temperature, the area-law entropy contribution to $\bar{I}_{l,l}$ dominates for small $l$, but the influence of the Page mutual information, which acts as a lower bound, is present at larger $l$. The Page mutual information grows linearly in $l$ after $l > L/4$, although we did not investigate this regime.

\section{Numerical Results: Exponential relaxation behavior at late times}%
\label{sec:exponential_relaxation}

\subsection{Introduction}%

One aspect of the thermalization of closed many body systems which has not been addressed numerically very thoroughly is the temporal behaviour of observables at finite, but long times. Our enhanced precision enabled by the reduced temporal fluctuations for large system sizes, allow us to study the finite size, long time behaviour in a systematic and rather accurate manner. We find clear evidence for exponential relaxation behavior in time and study the relaxation time dependence on various factors such as system size, energy density and variance and observables considered.

\subsection{Functional Form}%

\begin{figure}
	\centering
	\plot{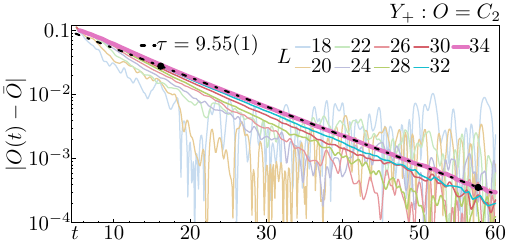}
	\put(-210, 110){(a)}

	\plot{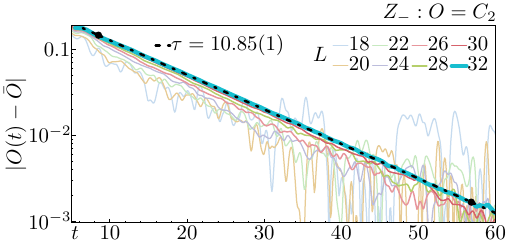}
	\put(-210, 110){(b)}
	\caption{Exponential decay of the energy correlator $C_2$ on a linear-log scale, with fitted exponentials shown with dashed lines. The two dots along the fitted lines mark the limits of the time interval where the fitting was performed.}%
	\label{fig:logO_t}
\end{figure}

Our late time numerical data shows an exponential relaxation of local operator expectation values at long times (Fig.~\ref{fig:logO_t}). The exponential decay is not detected across all simulated system sizes and states, but rather starts emerging as $L$ increases and the temporal fluctuations decrease exponentially (c.f.~Sec.~\ref{sub:fluctuations_around_equilibrium}), and due to this very fast decrease one can loosely identify a lower bound $L$ (or small range of sizes) in which the exponential relaxation is first apparent. This threshold system size is dependent on the operator and state but overall it is lowest for the energy correlators, and is lower for higher temperature and lower variance states.

We estimate a decay time scale $\tau$ ($e^{-t/\tau}$) by a linear fit on the absolute deviation $\log|O(t) - \bar{O}|$ on some time interval that was manually picked for each operator, state, and system size, as seen in Figs.~\ref{fig:logO_t} and~\ref{fig:logO_t_scaled}. This interval grows with $L$ as the temporal fluctuations reduce in size. For small $L$ the fitted $\tau$ can change significantly depending on the chosen interval. The error bars shown in some plots of this section do not take this into account as they are merely confidence intervals derived from the fitting. However, while the accuracy of individual data points could be challenged, collectively the data show trends consistent across different parameters which we deem robust enough to the arbitrariness in the choice of limits. Note also that the convergence displayed in Fig.~\ref{fig:logO_t} covers between two and three decimal orders of magnitude, so it is quite a compelling exponential decay.

\subsection{Dependence on system size, energy density, variance and operator considered}%

\begin{figure}
	\centering
	\plot{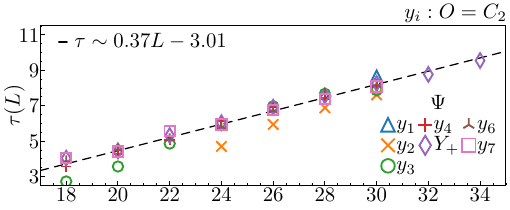}
	\put(-225, 92){(a)}

	\plot{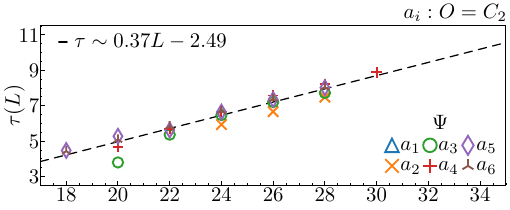}
	\put(-225, 92){(b)}

	\plot{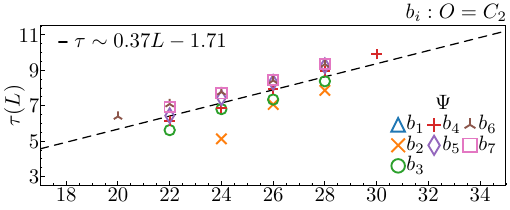}
	\put(-225, 92){(c)}

	\caption{System size dependence of the time scale $\tau$ for three equal energy series of states. The dashed line is a linear fit on interpolated data of $\tau$ at the same energy variance for each system size.}%
	\label{fig:tau_L}
\end{figure}

\begin{figure}
	\centering

	\plot{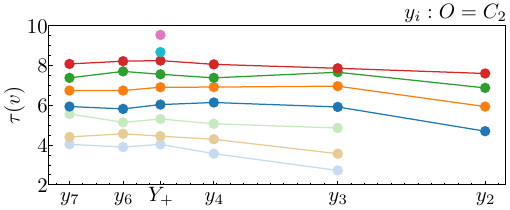}
	\put(-220, 92){(a)}

	\plot{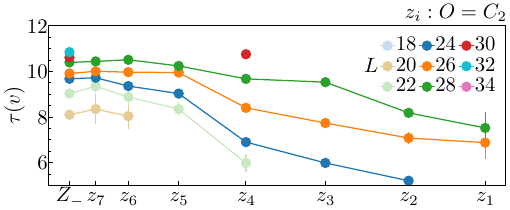}
	\put(-220, 92){(b)}

	\plot{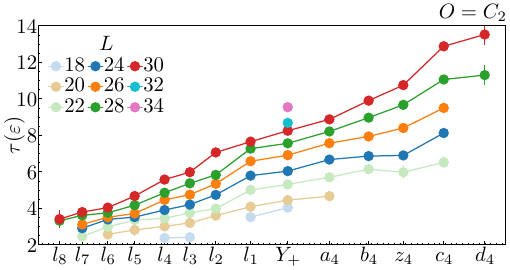}
	\put(-220, 121){(c)}

	\caption{The time scale $\tau$ as a function of the energy variance $v$ for the $y_i$ (a) and $z_i$ (b) series of states, and as a function of the energy density $\En$ for the $v = \tilde{v}$ and $l_i$ series of states (c). Note the weak dependence on $v$ and the approximately linear growth with $\En$.}%
	\label{fig:tau_ev}
\end{figure}

\begin{figure}
	\centering

	\plot{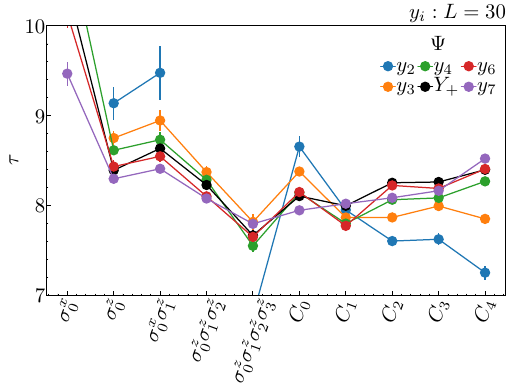}
	\put(-220, 179){(a)}

	\plot{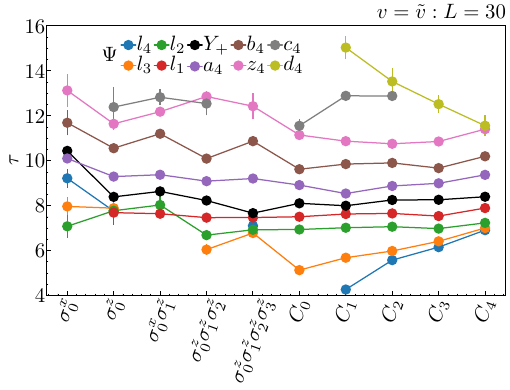}
	\put(-220, 179){(b)}

	\caption{The time scale $\tau$ for different local operators $\Op$ for the $y_i$ (a) and $v = \tilde{v}$ (b) series of states. The error bars are confidence intervals derived from the fitting. Overall, $\tau$ is weakly dependent on the operator.}%
	\label{fig:tau_O}
\end{figure}

We observe that the relaxation time $\tau$ grows linearly with system size (Fig.~\ref{fig:tau_L}), it depends approximately linearly on energy density (Fig.~\ref{fig:tau_ev}a), and depends only weakly on energy variance density but with a noticeable decrease/increase (depending on the operator) with variance at finite temperature (Fig.~\ref{fig:tau_ev}b,c). Also, $\tau$ is rather independent of the operator (Fig.~\ref{fig:tau_O}), with a stronger dependence appearing at lower temperatures.

\begin{figure}
	\centering
	\plot{logO-t,L,shifted,psi=Y_+,O=C2.pdf}
	\put(-231, 110){(a)}

	\plot{logO-t,L,shifted,psi=Z_-,O=C2.pdf}
	\put(-231, 110){(b)}
	\caption{Exponential decay of the energy correlator $C_2$ on a linear-log scale, with the fitted exponential shown with black dashed lines. These are the same data lines shown in Fig.~\ref{fig:logO_t}, except they have been shifted from each other for better visualization as they have a significant overlap ($L$ is increasing from bottom to top). Two black dots along the fitted lines mark the limits of the time interval where the fitting was performed.}%
	\label{fig:logO_t_scaled}
\end{figure}

While the fitted linear rates of growth of $\tau(L)$ in Fig.~\ref{fig:tau_L} seem to be close in the three cases, we are conservative in interpreting this result and we think this matter should be refined with more data. The offset of the fitted lines are increasing with $\En$. While an extrapolation to lower $L$ implies a negative $\tau$ at some finite size, observing any exponential decay at small system sizes is unrealistic, thus $\tau$ is ill-defined at smaller $L$ and extrapolation is not meaningful.

The growing dependence of $\tau$ on the energy variance as we move away from infinite temperature is more evident in Fig.~\ref{fig:tau_ev}a and b, where we see the decrease of $\tau$ at larger energy variance for the operator $\hat{C}_2$. The dependence of $\tau$ on $v$ is stronger for local spin operators, meaning we do not observe the clear linear dependencies with system size as we do for $\hat{C}_2$ and other energy correlators.

In Fig.~\ref{fig:tau_ev}c we plot the available fitted $\tau$ of the series of $v = \tilde{v}$ states (plus the remaining $l_i$ states) as a function of the energy density of those states, suggesting an approximately linear growth of $\tau$ with $\En$. This result is somewhat surprising to us, as one might have expected that the infinite temperature case (corresponding to $\En=0$) would have shortest relaxation time, while the limit of $\En$ going to the negative energy ground state could perhaps be expected to exhibit critical slowing down, i.e.~a significant enhancement of the relaxation time to the closeness to a quantum critical point in the phase diagram. Our numerical observations clearly point to the need to further investigate the relaxation behavior of finite size and infinite systems more thoroughly.

We also extracted decay time scales $\tau$ for certain Pauli spin operators and compare them together with short-range energy correlators in Fig.~\ref{fig:tau_O}. At infinite temperature (Fig.~\ref{fig:tau_O}a), the decay time scales across all operators (excluding $\hat{\sigma}^x$) are of comparable magnitude, except for $y_2$ which shows a large variability in time scales and also the largest uncertainty, with some fitted $\tau$ being excluded for this state as they were either deemed too inaccurate or no reasonable interval to fit on was found. The same inaccuracy is observed for $y_1$ on all operators, a fact which is common for most high variance states we simulated.

Curiously, the variance dependence of $\tau$ is operator dependent, as we can observe consistently in Fig.~\ref{fig:tau_O}a. For example, $\tau$ mostly decreases with $v$ for $\hat{C}_2$, while the opposite is observed for $\hat{C}_0$ or $\hat{\sigma}^z$.

The operator dependence is more significant at lower (absolute) temperatures (Fig.~\ref{fig:tau_O}b), supposedly due to an increased importance of the particular physics of the chosen model. Of all the measured local operators, $\hat{\sigma}^x$ stands out as being less correlated to the other operators in terms of the time scale. Coincidentally or not, $\hat{\sigma}_x$ is the only measured operator which does not explicitly appear in $\Ha^2$.

In Appendix~\ref{sec:relaxation_of_entanglement_entropy} we briefly discuss the dynamics of the entanglement entropy, where we also observed an exponential relaxation in time.

\section{Interpretation of the Exponential Relaxation Behavior at late times:
  Ballistic versus Diffusive Thermalisation}

The observation of exponential relaxation towards the steady state and the
proportionality of the relaxation time $\tau \sim L$ is unexpected at this stage.
In systems without any conserved quantities one would often expect an exponential relaxation with a thermalization time which does not depend on system size~\cite{Klobas2021,Mori2024,Jacoby2024,Zhang2024} Dual unitary cirquits are a good illustration of this behavior.
On the other hand in systems with conserved quantities, such as energy (the case under consideration here), magnetization or charge, it has been advocated that the relaxation of observables overlapping with conserved quantities is governed by fluctuating hydrodynamics~\cite{Lux.Muller.ea:2014,Wienand:2024}, with the expectation that the relaxation displays a power-law in time, and in 1+1D diffusive systems in particular exhibits a $1/\sqrt{t}$ hydrodynamical tail.

\subsection{Finite size effects in diffusion}%

At this stage it is important however to remind ourselves that in our study we are confined to a finite system of length $L$ and the long time behavior even for ordinary classical diffusion differs between finite size and infinite systems. In a finite size system governed by the classical diffusion equation, the longest wavelength Fourier mode of the conserved density decays with an exponential relaxation time $\tau_D~\sim~L^2/D$, with $D$ the diffusion constant, i.e.~for $t\gtrsim \tau_D$ the decay is $\sim \exp(-t/\tau_D)$. For shorter times $ O(1) \lesssim t \lesssim \tau_D$ one expects to observe a $1/\sqrt{t}$ power law tail in the relaxation of {\em local} densities as expected for the infinite system. As the system size $L$ increases, the power-law tail extends to later times scaling as $L^2$. The lower limit $O(1)\lesssim t $ denotes a microscopically controlled time scale required to establish the hydrodynamic regime.

\subsection{Probing diffusion in a finite size setup}%

Since our considered quantum many body system is chaotic and conserves energy, a natural
expectation is that it is indeed diffusive, and one can wonder whether we can observe diffusion on our accessible system sizes using ED. Then we can also address whether the observed relation time in Sec.~\ref{sec:exponential_relaxation} for a translation invariant situation is distinct from the Thouless time $\tau_D$ on the same system size due to diffusion.

\begin{figure}
	\centering
	\includegraphics[width=0.5\linewidth]{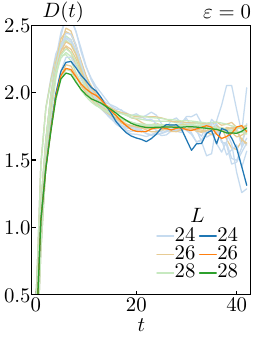}%
	\put(-107, 0){(a)}
	\includegraphics[width=0.5\linewidth]{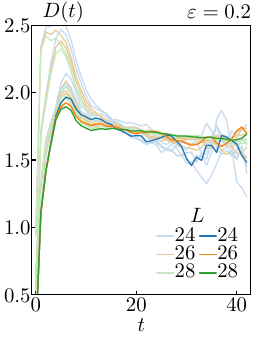}
	\put(-107, 0){(b)}
	\caption{Diffusion constant $D(t)$ (Eq.~\ref{eq:diffusionconstant}). For each energy density $\varepsilon$ we consider $7$ different domain-wall initial states with $\delta \varepsilon = 1$ and we plot the measured diffusion constant for $L=24,26,28$. We highlight with a stronger color the results of one state per $\varepsilon$ to better visualise the diminishing finite-size effects.}%
	\label{fig:DiffusionConstant}
\end{figure}

In order to address this question we use a different type of initial states for the following set of simulations. We prepare a system with periodic boundaries in a product state with $L/2-1$ consecutive bonds with energy expectation values $\En-\delta \En$, followed by one bond at energy $\En$ and then $L/2-1$ bonds with energy expectation values $\En+\delta \En$. The last missing bond is then again at mean energy $\En$. We find multiple (random) product state solutions for a range of $\En$ and $\delta \En$ by numerically optimizing the identical angles of the spins on the first  and second set of $L/2$ spins to agree with the required energy pattern on the bonds. This initial state has a domain wall profile for the conserved energy, and we expect the longest wavelength component of the energy density
\begin{equation*}
	\delta \varepsilon(q,t)\propto \sum_j \exp(-i q r_j) \langle\Ha_j\rangle_t
\end{equation*}
to decay with a relaxation time
\begin{equation}
	\tau_D^{(L)}(t)=- \frac{1}{\frac{d}{dt} \ln |\delta \varepsilon(q{=}2\pi/L,t)|} =\frac{L^2}{4 \pi^2 D(t)}\ .
	\label{eq:diffusionconstant}
\end{equation}
We observe a Thouless time of about $\tau_D\approx 12$ for $\En=0$ and $L=28$, while the local observables decay with a relaxation time of $\tau\approx 9$ for the same energy density and system size.
We can then convert the Thouless times into an estimate for the energy diffusion constant using Eq.~\eqref{eq:diffusionconstant} for different mean energies $\En$. The results in Fig.~\ref{fig:DiffusionConstant} show a (running) diffusion constant for $\En=0$ (left panel) trending towards $D\approx 1.7(1)$ for longer times between 20 and 50. For even larger times the considered Fourier component of the energy is zero within the residual fluctuations and the extracted diffusion constant becomes noisy and blows up. The fact that the running diffusion constants for the two largest system sizes ($L=26$ and $L=28$) lie on top of each other for a considerable time window highlights that the underlying relaxation times indeed diverge as $\tau \sim L^2$ in the diffusion setup in contrast to $\tau \propto L$ as observed for the translation invariant situation in Sec.~\ref{sec:exponential_relaxation}.

The value $D\approx 1.7(1)$ for the diffusion constant is in good agreement with a recent information lattice based method based study~\cite{Klein-Kvorning.Herviou.ea:2022} but is quite a bit smaller
than one reported in an older work~\cite{Leviatan.Pollmann.ea:2017} using matrix product states with a small bond dimensions at long times. The value is also in good agreement with a light-cone MPS study~\cite{Frias-Perez:2022}.

We can also explore the dependence of the diffusion constants as a function of energy and we observe that the running diffusion constants show a decreasing trend when increasing the energy density from $\En=0$ (left panel) to $\En=0.2$ (right panel). We will discuss a possible explanation for this trend below, but also note that for $\En=-0.2$ the running diffusion constant has not fully converged among different initial states in the available time window. It is also remarkable that overall our running diffusion constants still decrease with time up to the loss in precision. At this stage it is not clear how accurate the ED results are because of the relatively small system sizes and the different setup (we study finite-size, long time effects in a thermalizing post-quench setup, while the tensor network studies often  track local density inhomogeneities spreading in space and time for large systems initially in thermal equilibrium). However since ED performs no truncation of the Hilbert space or operator sizes, there is a genuine possibility that ED detects effects beyond the truncation schemes of tensor network or other entanglement based schemes. The fact that the initial study of Ref.~\cite{Leviatan.Pollmann.ea:2017} with a small bond dimension reported
a diffusion constant of $D\approx2.2$ and our data for different initial states also reach such high values at short times could be taken as mild support for such a scenario.

\subsection{Ballistic thermalisation of equal-time correlation functions}%

After this excursion into a diffusion study of Hamiltonian~\eqref{eq:BCH}
we can conclude that the relaxation times $\tau \sim L$ observed for equal-time
correlators are not of diffusion origin. While we could only numerically
determine the Thouless time up to $L=28$ in the diffusion setup, our extrapolation based on the extracted diffusion constant to system size $L=34$ would yield a Thouless time of about $17$, while the relaxation times in the translation invariant case are only found to be about $10$ for the same system size $L=34$ and energy density $\En=0$.

This raises the question of the origin of the faster relaxation times in the translation invariant case. One candidate theory for the relaxation of equal-time correlations is
fluctuating hydrodynamics. The idea is that correlations can relax and reach their long
time values primarily through diffusive processes in the post-quench system~\cite{Lux.Muller.ea:2014,Wienand:2024}. On the other hand diffusion is not the only propagating {\em information} carrying mode in an ergodic system, even though it
is the sole mode which can equilibrate actual density inhomogeneities. The main
idea we put forward here is that the equilibration of equal-time correlation functions can also be achieved by ballistic modes which are underlying the ballistic OTOC and scrambling dynamics observed in many systems. The fact that correlation functions can relax by ballistic modes has already been observed in earlier work in the context of Lieb-Robinson light cones~\cite{Calabrese2006,Lauchli.Kollath:2008,Cheneau2012}.

\begin{figure}
	\centering
	\plot{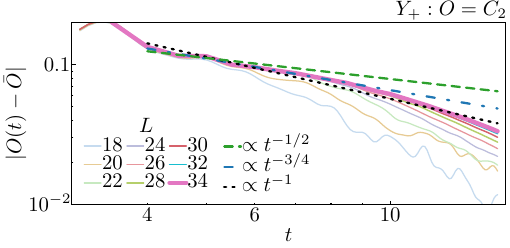}
	\caption{The decay of $\hat{C}_2$ for the $Y_+$ state on a log-log scale. The fluctuations at times earlier than $4$ are much larger.}%
	\label{fig:logO_logt}
\end{figure}

At this stage the question of whether the observed $\tau\sim L$ behavior translates into a power law in time, e.g. $1/t$ is difficult to answer based on the available data, see Fig.~\ref{fig:logO_logt}.  It is difficult to see a power law appearing between $O(1)\lesssim t \lesssim \tau(L)$, because $\tau(L) \sim 10$.

We do not identify a power-law regime of relaxation. The energy correlators follow the sum rule $\sum_r \braket{\hat{C}_r(t)} = v$ and thus we could expect the power-law decay of $\hat{C}_r$, if present, to be diffusive ($t^{-1/2}$)~\cite{Lux.Muller.ea:2014}. If we interpret the observed exponential decay as an exponential cutoff of a power-law decay at finite sizes, then we should find such a decay in some interval before the exponential regime. As seen in Fig.~\ref{fig:logO_t}a, for $Y_+$ there is an interval at early times and larger system sizes (before the marked fitting limits) where the decay does not exactly agree with an exponential. In Fig.~\ref{fig:logO_logt} we plot this interval on a log-log scale. From the comparison of the data to a power-law, it seems more likely to us that this is a feature of the time evolution dependent on the operator and state, and not the emergence of a power-law regime, which in any case would be closer to a $\sim t^{-3/4}$ decay than diffusive.


\section{Conclusions}%
\label{sec:conclusions}

We systematically studied the finite-size deviations and relaxation time scales of local operators and entanglement entropy, and extensively covered their dependence on the system size, energy density, and in particular we highlight the energy variance dependence, which we believe is often overlooked.

Simulating sufficiently large system sizes proved to be critical to both access the regime where the ETH scaling~(\ref{eq:ETH_scaling}) becomes sufficiently accurate, and to properly study the dynamics: the exponential decay of the fluctuations as system size is increased causes a sharp transition between noisy dynamics and a clear exponential relaxation towards the diagonal ensemble (Figs.~\ref{fig:logO_t} and~\ref{fig:logO_t_scaled}).

The observed scaling of the finite-size deviations of local operators agrees perfectly with the ETH prediction~(\ref{eq:ETH_scaling}). The observed quadratic scaling of the entropy deviations is a novel result, but which has in the meantime been explained analytically within the ETH framework~\cite{Huang:2025}.

The growth of an area-law term at finite temperature in the entanglement entropy is consistent with a non-zero mutual information from the presence of correlations between spin clusters, and we observed a clear agreement between the quantum and thermal counterparts of the mutual information for small clusters. The mutual information of finite temperature states is dominated by the area-law term for small cluster sizes but is supplanted by the Page mutual information at larger cluster sizes.

An exponential relaxation is accurately observed and is described by a single time scale $\tau$ which grows linearly with system size, approximately linearly with energy density, and depends weakly on the energy variance and local operator in question. It would be interesting to explore whether the relaxation time spectrum and its finite-size behaviour reveals quantitative aspects of the long-time, long-wavelength field theory description, as is well known in the context of finite size spectra and e.g.~conformal field theories~\cite{Cardy:1984}.

The present work opens up several avenues for future research: Several scaling dependencies were established but not all are fully understood in terms of their physical origin. This includes the observed quadratic scaling of the entropy deviations and the linear growth of the relaxation time scale with system size and energy density.

It would be interesting to contrast the current results with other models, and to study the dynamics of non-uniform states, which should provide further insight into the physical origin of the observations.

\acknowledgments

We acknowledge helpful discussions with D.~Abanin, V.~Alba, C.~Aron, M.C.~Banuls,  A.~Belin, A.~Elben, M.~Haque, L.~Herviou, Y.~Huang, A.~Michailidis, S. Pappalardo, A. Rosch, and R.~Steinigeweg.
The computational results have been obtained on the Merlin6 cluster at PSI and the Fiamma LLTCP IPHYS server.

\appendix

\section{ETH predictions}%
\label{sec:ETH_prediction}

For eigenstates and for typical quantum states of energy $E$ and energy variance scaling linearly with the system size $L$, the diagonal ensemble expectation value which we estimate from $\bar{O}$ is expected to agree with the expectation value in the microcanonical ensemble at energy $E$ in the $L \rightarrow \infty$ limit if the system is non-integrable. Since the micro and canonical ensembles approach each other in the thermodynamic limit, the same can be said for such an ensemble at the inverse temperature $\beta$ that results in an average thermal energy $E$.

This agreement is formulated more precisely for finite $L$ in the ETH~\cite{Deutsch:1991, Srednicki:1994, Rigol.Dunjko.ea:2008}.

ETH states that the matrix elements of any compact quantum operator between eigenstates of a non-integrable Hamiltonian are diagonal up to corrections exponentially small on the system size, where the diagonal elements are purely a function of the eigenstate energy.

Take $\ket{\En}$ to be an eigenstate of energy density $\En$. Following the ETH, we write the diagonal matrix element of such an operator as
\begin{equation}
	\braket{\En|\Op|\En} \approx O(\En) + \mathcal{O}(\Omega^{-1/2}(\En)),
\end{equation}
for some operator-dependent smooth function $O(\En)$, and where $\Omega(\En)$ is the density of states which scales exponentially with the (thermal) entropy $\tilde{S}$: $\Omega(\En) \sim 2^{\tilde{S}} \sim 2^{rL}$ for some energy dependent rate $r \leq 1$, where the entropy is written in units of $\log_2$.

Given that for large enough system sizes the diagonal matrix elements are approximately a function of the energy only, it suffices to know the probability that the system has an energy density $\En$ (which we define as the probability density $p(\En)$) to calculate all expectation values from the functions $O$:
\begin{equation}
	\braket{\Op}_\K \approx \int p_\K(\En') O(\En') d\En',
	\label{eq:O_ETH}
\end{equation}
where $\K$ represents a statistical ensemble. Whether $\K$ is a microcanonical, canonical ($\CE$), or a quantum diagonal ensemble ($\DE$), Eq.~(\ref{eq:O_ETH}) is applicable, at least on some mid-spectrum region.

For an eigenstate or for the microcanonical ensemble, $p(\En)$ is a Dirac distribution. For the canonical ensemble, it is known that $p(\En)$ obeys a central limit theorem and thus it goes asymptotically to a normal distribution of energy variance $\tilde{v}/L$, where $\tilde{v}$ is system size independent. Similarly, the energy distribution of product states also converges towards a normal distribution~\cite{Hartmann..ea:2004}. Thus, if both the energy densities and variances of the diagonal and canonical ensembles agree, we have
\begin{equation}
	\bar{O} \approx \tilde{O}.
\end{equation}

How do these two quantities differ when the variances are not equal? Expanding $O(\En')$ as a Taylor series around $\En$ in Eq.~(\ref{eq:O_ETH}) and taking $p(\En') = \mathrm{N}(\En, v_\K/L)$ to be normal, we obtain:
\begin{align}
	\braket{\Op}_\K & \approx O(\En) + \sum_{n=1}^{\infty} \frac{1}{2n!!}{\left( \frac{v_\K}{2L} \right)}^{n} \partial^{2n}_{\En} O(\En), \\
	\label{eq:ETH_second_order}
	                & \approx O(\En) + \frac{v_\K}{2L} \DD O(\En) + \frac{v^2_\K}{8L^2} \partial^4_{\En}O(\En) + \dots,
\end{align}
Eq.~(\ref{eq:ETH_scaling}) then follows. Note that considering the fourth order terms gives a term proportional to $v^2-\tilde{v}^2$ in the deviations, which is not related to the quadratic dependence observed for the difference between entropies in Sec.~\ref{sec:thermalization_of_entanglement_entropy} where instead we see an approximate dependence of the form ${(v-\tilde{v})}^2/L^2$.

\section{Conservation of energy variance}%
\label{sec:conservation_of_energy_variance}

\begin{figure}
	\centering
	\plot{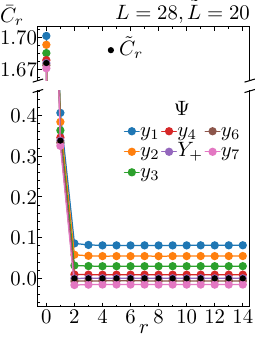}%
	\put(-106, 153){(a)}
	\plot{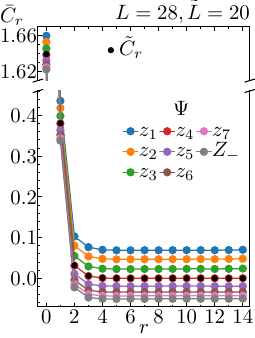}
	\put(-106, 153){(b)}
	\caption{Equilibrium average $\bar{C}_r$ of the energy correlators as a function of $r$ for two equal energy series of states. The thermal variance states ($Y_+$ and $z_4$) coincide well with the thermal average shown with black dots.}%
	\label{fig:C_r}
\end{figure}

Due to translation invariance, we have
\begin{equation}
	v = \sum_r \braket{\Psi|\hat{C}_r(t)|\Psi},\quad \tilde{v} = \sum_r \tilde{C}_r.
	\label{eq:variance_conservation}
\end{equation}

The effects of the sum rules~(\ref{eq:variance_conservation}) on the energy correlators are seen in Fig.~\ref{fig:C_r}: The energy correlators are all larger/smaller than the thermal counterparts depending on whether the states' energy variance density is higher/lower than the thermal.

\begin{figure}
	\centering
	\plot{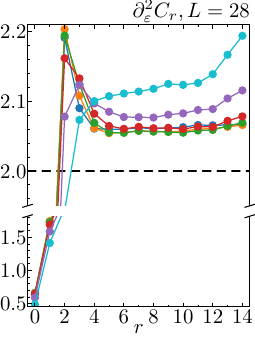}%
	\put(-110, 154){(a)}
	\plot{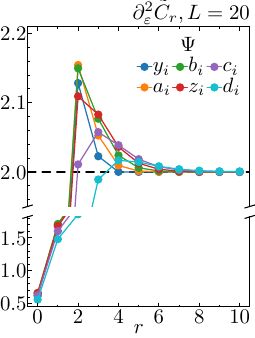}
	\put(-110, 154){(b)}
	\caption{Estimates of the ETH prefactor $\DD O$ for the energy correlators. (a) Estimates from linear fits using data from equal energy series of states. (b) Second derivative of the thermal expectation.}%
	\label{fig:delta_C_r}
\end{figure}

For each $r$, the spread of the energy correlators across all states considered is smaller for $r=0,1$, while it is larger but approximately constant in $r$ for $r>1$. As seen in Fig.~\ref{fig:delta_C_r}, this follows from the fact that the prefactors $\DD C_r$ are smaller for $r=0,1$ than for $r>1$ where they are in fact approximately constant ($\DD C_r \gtrsim 2$).

The proximity of larger values of $\DD C_r$ to $2$ follows from the independence between energy operators beyond the correlation length:

\begin{equation}
	\lim_{r \to \infty} \DD C_r = 2,
	\label{eq:limit_of_c_r}
\end{equation}
as derived in Appendix~\ref{sec:derivatives_of_o_en}.

\begin{figure}
	\centering
	\plot{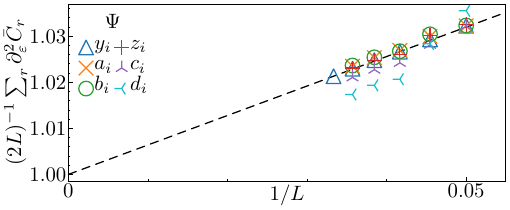}
	\plot{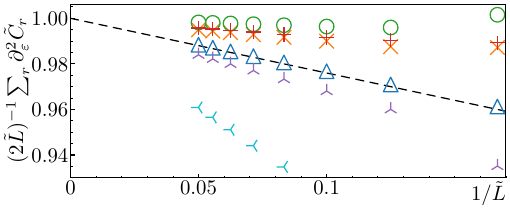}
	\caption{The sum of the ETH prefactor estimates for the energy correlators, divided by the expectation~(\ref{eq:sum_rule_derivatives}). The dashed lines are visual guides that cross $(0, 1)$.}%
	\label{fig:sum_delta_C_r}
\end{figure}

Moreover, the sum rules~(\ref{eq:variance_conservation}) force similar sum rules on the prefactors $\DD C_r$: Starting from the ETH scaling supposition (Eq.~(\ref{eq:ETH_scaling})) for the $C_r$ operators, summing over $r$ and using Eq.~(\ref{eq:variance_conservation}), we obtain
\begin{equation}
	2L \sim \sum_{r} \DD C_r(\En) \approx \sum_{r} \DD \tilde{C}_r(\En),
	\label{eq:sum_rule_derivatives}
\end{equation}
which should hold asymptotically as $L \to \infty$. We verify this in Fig.~\ref{fig:sum_delta_C_r}. Note that the diagonal and thermal estimates reach this limit differently, as exemplified in Fig.~\ref{fig:delta_C_r}: In the diagonal case, the sum is slightly above $2L$ and this excess is spread somewhat equally across the derivatives of non-local correlators, thus the excess plateau decreases uniformly as $L$ increases, while in the thermal case the second derivative decays quickly towards $2$ at higher $r$, thus the limit of $2L$ is reached simply by the addition to the sum of more terms equal to $2$ as $\tilde{L}$ increases.

\section{Estimation of derivatives of $O(\En)$}%
\label{sec:derivatives_of_o_en}

In this appendix we collect expressions related to the derivatives of $O(\En)$. The expectation values that follow are taken in the thermal ensemble ($\braket{\Op} = \tilde{O}$). The term $\DD O(\En)$ can be estimated from the thermal ensemble if one has full spectrum data:
\begin{equation}
	\begin{aligned}
		\DD O(\En) \approx & \DD\tilde{O}(\En)                                                                                                                                   \\
		=                  & \left( \frac{L}{\tilde{v}} \right)^{3}\left[\braket{\ha^3} - \En^3\right]\left[ \En\braket{\Op} - \braket{\ha \Op} \right]                          \\
		+                  & \left( \frac{L}{\tilde{v}} \right)^{2}\left[ \braket{\ha^2\Op} + \En\braket{\ha\Op} - 2\En^2\braket{\Op} \right] - \frac{L}{\tilde{v}}\braket{\Op}.
	\end{aligned}
	\label{eq:derivatives_of_o_en}
\end{equation}
where we defined $\ha \equiv \Ha/L$ \footnote{Note that Eq.~\ref{eq:derivatives_of_o_en} seems to be equivalent to Eq.~48 in~\cite{Huang:2024}}. This follows from
\begin{equation}
	\begin{aligned}
		\ddv{\braket{\Op}}{\En} & = \frac{\partial}{\partial \En} \left( \dv{\beta}{\En} \dv{\braket{\Op}}{\beta} \right)                      \\
		                        & = \ddv{\beta}{\En} \dv{\braket{\Op}}{\beta} + \left( \dv{\En}{\beta} \right)^{-2} \ddv{\braket{\Op}}{\beta},
	\end{aligned}
\end{equation}
where
\begin{align}
	\ddv{\beta}{\En} & = - \ddv{\En}{\beta}\left( \dv{\En}{\beta} \right)^{-3},          \\
	\dv{\En}{\beta}  & = L\left[ \En^2-\braket{\ha^2} \right] = -\tilde{v},              \\
	\ddv{\En}{\beta} & = L^2\left[ \braket{\ha^3} - 3\En\braket{\ha^2} + 2\En^3 \right],
\end{align}
and
\begin{align}
	\dv{\braket{\Op}}{\beta}  & = L\left[ \En\braket{\Op} - \braket{\ha \Op} \right] ,                                              \\
	\ddv{\braket{\Op}}{\beta} & = L^2\left[ \braket{\ha^2\Op} - 2\En\braket{\ha \Op} +(2\En^2 - \braket{\ha^2})\braket{\Op} \right]
\end{align}
Then Eq.~(\ref{eq:derivatives_of_o_en}) follows.

This quantity simplifies when $\Op = \hat{C}_r$ for $r > \xi$ (the correlation length). We have
\begin{equation}
	\braket{\hat{C}_r} = \braket{\Ha_0\Ha_r} - \En^2 \rarrow \braket{\Ha_0}\braket{\Ha_r} - \En^2 = 0.
\end{equation}
\begin{equation}
	\begin{aligned}
		\braket{\ha \hat{C}_r} = & \frac{1}{L}\sum_{i} \braket{ \Ha_i(\Ha_0 - \En)(\Ha_r - \En) }                   \\
		\rarrow                  & \frac{2}{L} \sum_{i \ll r} \braket{ \Ha_i(\Ha_0 - \En)}\braket{\Ha_r - \En} = 0.
	\end{aligned}
\end{equation}
\begin{equation}
	\begin{aligned}
		\braket{\ha^2 \hat{C}_r} = & \frac{1}{L^2}\sum_{i,j} \braket{ \Ha_i\Ha_j(\Ha_0 - \En)(\Ha_r - \En) }                                              \\
		\rarrow                    & \frac{2}{L^2} \left[ \sum_{i \ll r} \braket{ \Ha_i(\Ha_0 - \En)} \right]^2 = 2 \left( \frac{\tilde{v}}{L} \right)^2.
	\end{aligned}
\end{equation}
Finally, by substituting in Eq.~(\ref{eq:derivatives_of_o_en}) we get Eq.~(\ref{eq:limit_of_c_r}).

\section{Entropy of $Y_+$}%
\label{sec:entropy_of_y}

\begin{figure}
	\centering
	\plot{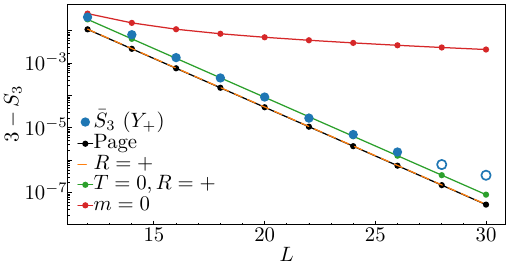}
	\caption{Difference of entanglement entropies on three site clusters with respect to the maximum $l=3$ on a log scale. We show the expected entropy of random states drawn from the full Hilbert space (Page, Eq.~(\ref{eq:page_entropy})) and reflection, translation and magnetization symmetry sectors. The two unfilled data points at $L=28, 30$ were affected by accumulated numerical error during the calculation of the reduced density matrix of $Y_+$.}%
	\label{fig:page_entropy}
\end{figure}

The $Y_+$ state is the only infinite temperature state we consider whose variance density coincides with the thermal one, and if we assume that higher moments of energy are negligible, then we can expect $Y_+$ to be equivalent to a random state. The expected entanglement entropy of such a state is known as the Page entropy (Eq.~(\ref{eq:page_entropy})). However, the state $Y_+$ is in the reflection symmetric and zero momentum symmetry sector, and the entropy of random states drawn from such a subspace is further limited~\cite{Haque.McClarty:2022}. Different symmetries can show quite different leading corrections to the maximum $S_l=l$: For example, the entropy deviation on a translation symmetry sector $T$ is still exponentially small in $L$ like the Page entropy~\cite{Nakata.Murao:2020}, while a system with fixed magnetization $m$ or particle conservation results in a $\sqrt{L}$ deviation~\cite{Vidmar.Rigol:2017, Bianchi.Hackl.ea:2022}.

To our knowledge, the average entropy on the symmetry sector of our initial states ($T=0, R=+$) is not available in the current literature, so we estimated this quantity numerically by sampling random states on the sector~\footnote{Random quantum states are generated from random complex unitary transformations of an arbitrary normalized state of the Hilbert space~\cite{Ozols:2009}. However, generating large unitary matrices is not feasible for the system sizes we consider, even when we restrict ourselves to a symmetry sector of interest. An approximate way we implement is to take random vectors with complex entries $z \propto r e^{i \phi}$ where $r$ is normal, $\phi$ is uniformly distributed and the proportionality constant is given by normalization. This makes the numerical calculation of the average entanglement entropy feasible for $L \sim 30$, which we have carried out.}.
We repeated the numerical procedure for the reflection positive symmetry sector ($R=+$), while we know the result exactly for the zero magnetization sector ($m=0$)~\cite{Vidmar.Rigol:2017, Bianchi.Hackl.ea:2022}. We plot these quantities for $l=3$ in Fig.~\ref{fig:page_entropy}. The equilibrium entropy of the $Y_+$ state agrees with the entropy of a random state from the symmetry sector the state belongs to. The disagreement at $L=28, 30$ is due to accumulated numerical error while explicitly calculating the reduced density matrix of $Y_+$.

\section{Heisenberg time}%
\label{sec:heisenberg_time}

Since no exponential decay is detectable once the expectation decays to values below the fluctuations, we could consider that equilibrium has been reached at this point, at least in terms of local observables, so $\tau$ could be an appropriate time scale to define the thermalization of such variables, although full relaxation is not achieved at least until the exponentially large Heisenberg time.

\begin{figure}
	\centering
	\plot{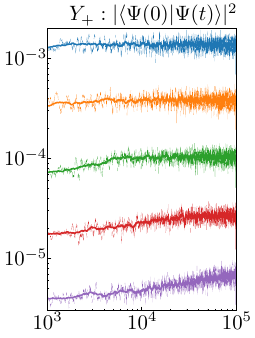}%
	\put(-110, 157){(a)}
	\plot{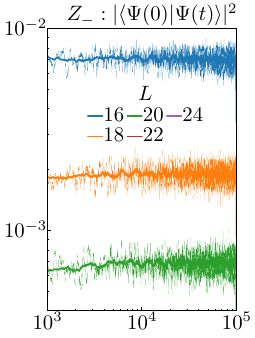}
	\put(-110, 157){(b)}
	\caption{The return probability (fidelity) over time on a log-log scale, averaged over 1000 (thick line) and 100 (thin dashed line) time units, for the $Y_+$ and $Z_-$ initial states.}%
	\label{fig:Heisenberg}
\end{figure}

We calculated the fidelity $|\braket{\Psi(0)|\Psi(t)}|^2$ over time for two states, up to a final simulation time of $t_f = 10^5$, as seen in Fig.~\ref{fig:Heisenberg}. At such a large time scale, the fidelity seems to oscillate very rapidly, so we time-averaged the data to smooth out the oscillations. The fidelity fully relaxes to a plateau at an exponentially increasing time, the Heisenberg time, associated to the mean energy level spacing of the model. A ramp behavior before the plateau is observable at larger system sizes for the $Y_+$ state (Fig.~\ref{fig:Heisenberg}a). 

\section{Relaxation of entanglement entropy}%
\label{sec:relaxation_of_entanglement_entropy}

\begin{figure}
	\centering
	\plot{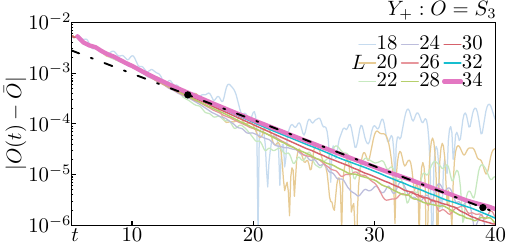}
	\put(-210, 110){(a)}

	\plot{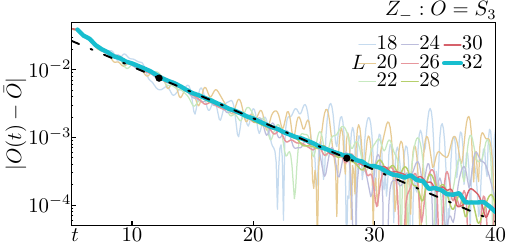}
	\put(-210, 110){(b)}
	\caption{Exponential decay of the three site entanglement entropy on a linear-log scale, with fitted exponentials shown with black dashed lines. Two black dots along the fitted lines mark the limits of the time interval where the fitting was performed.}%
	\label{fig:logS_t}
\end{figure}

We also observe a clear exponential decay of the entanglement entropy of small clusters (Fig.~\ref{fig:logS_t}), with associated decay times.

The numerical error that our method of calculating $S_l$ entails might be limiting the size of the observable exponential regime, especially for $S_3$ where the error is larger.

\begin{figure}
	\centering
	\plot{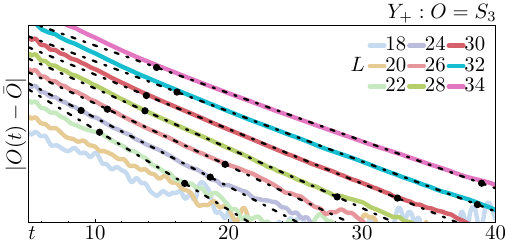}
	\put(-231, 110){(a)}

	\plot{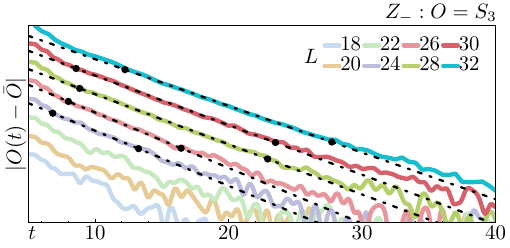}
	\put(-231, 110){(b)}
	\caption{Exponential decay of the three site entanglement entropy on a linear-log scale, with fitted exponentials shown with black dashed lines. These are the same data lines shown in Fig.~\ref{fig:logS_t}, except they have been shifted from each other for better visualization as they have a significant overlap ($L$ is increasing from bottom to top). Two black dots along the fitted lines mark the limits of the time interval where the fitting was performed.}%
	\label{fig:logS_scaled_t}
\end{figure}

\section{Full spectrum analysis}%

\begin{figure}
	\centering
	\plot{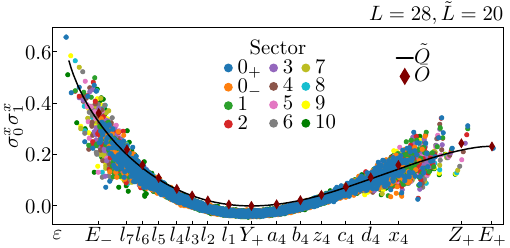}
	\put(-220, 109){(a)}

	\plot{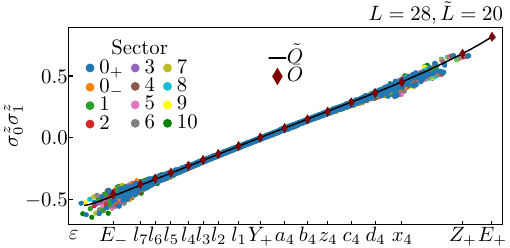}
	\put(-213, 109){(b)}

	\plot{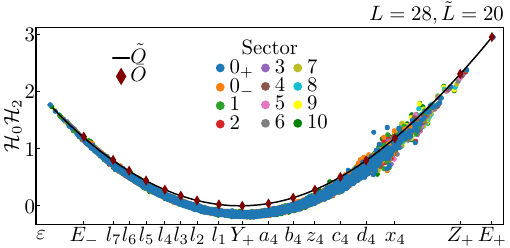}
	\put(-230, 109){(c)}
	\caption{The eigenvalues of several local operators for all eigenstates of the $\tilde{L}=20$ Hamiltonian per momentum symmetry sector. The thermal expectation curve is shown with a black line, while the diagonal ensemble estimate of certain states is shown with dark red diamonds. The curvature (second derivative) of the lines/surfaces determines the response of the diagonal/thermal deviations to the energy variance (Eq.~\ref{eq:ETH_scaling}).}%
	\label{fig:full_spectrum}
\end{figure}

In Fig.~\ref{fig:full_spectrum} we plot some local operator eigenvalues of all eigenstates of the $\tilde{L} = 20$ Hamiltonian, showing the thermal expectation curve and the diagonal ensemble estimate of selected states.

We can predict from the plots in Fig.~\ref{fig:full_spectrum} how the deviations between thermal and diagonal expectation values respond to changes in energy variance of the initial Bloch state by looking at the curvature of the eigenvalue curves/surfaces (Eq.~(\ref{eq:ETH_scaling})): The flatness of the $\sigma^z_0\sigma^z_1$ curve causes the smallest deviations, while the large curvature of $\Ha_0\Ha_2$ (and therefore $C_2$) causes the largest deviations.

\section{Bloch states}%
\label{sec:bloch_states}

Our initial product states are defined as
\begin{equation}
	\ket{\psi} = \bigotimes_j \left[ \cos\frac{\theta}{2}\ket{\uparrow}_j + e^{i\phi}\sin\frac{\theta}{2}\ket{\downarrow}_j \right],
\end{equation}
where the tensor product runs over all sites $j$ of the ring. The energy density of the initial states is
\begin{equation}
	\En \equiv \braket{\Ha_j} = \cos^2\theta + h_z\cos\theta + h_x\cos\phi\sin\theta.
\end{equation}
The initial expectation values of the energy correlators are
\begin{equation}
	\braket{\hat{C}_r} =
	\begin{cases}
		u_{xx}^2 + \frac{1}{2}u_x^2 + \frac{1}{2}u_y^2, & r=0,   \\
		\frac{1}{4}u_x^2 + \frac{1}{4}u_y^2,            & |r|=1, \\
		0,                                              & |r|>1,
	\end{cases}
	\label{eq:correlations}
\end{equation}
and the energy variance density is
\begin{equation}
	v \equiv \braket{\Ha^2}/L - L\En^2 = \sum_r \braket{\hat{C}_r} = u_{xx}^2 + u_x^2 + u_y^2,
\end{equation}
where we defined
\begin{align}
	u_{xx} & \equiv \sin^2\theta,                                                    \\
	u_x    & \equiv -2\cos\theta\sin\theta + h_x \cos\phi\cos\theta - h_z\sin\theta, \\
	u_y    & \equiv h_x\sin\phi.
\end{align}

\begin{figure}
	\centering
	\plot{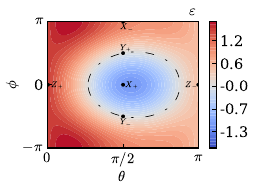}%
	\put(-97, 83){(a)}
	\plot{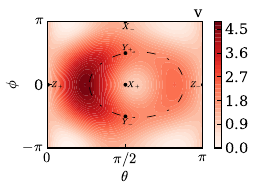}
	\put(-97, 83){(b)}

	\plot{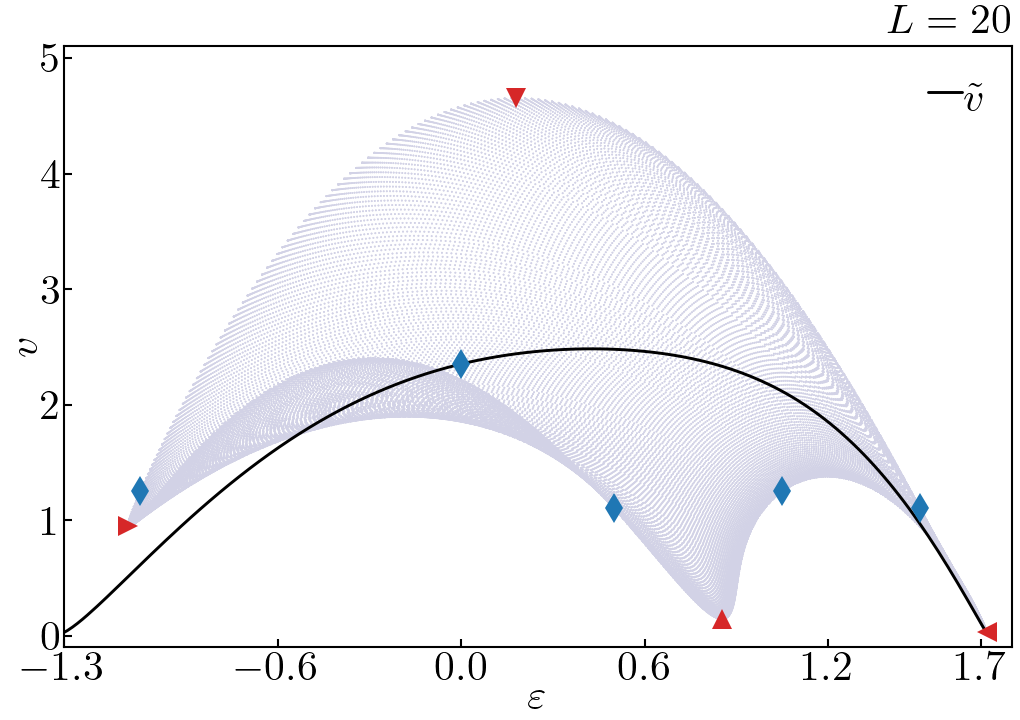}
	\put(-230, 166){(c)}
	\caption{Energy (a) and variance (b) per spin on the Bloch sphere $(\theta, \phi)$. The dashed oval is the $\En = 0$ line. (c) A uniform point density on the Bloch sphere is mapped on to the $(\En, v)$ surface, showing the density of available initial states. The surface is at least doubly degenerate due to the $y \leftrightarrow -y$ symmetry, with a small region between $E_-$ and $Z_-$ being four times degenerate.}%
	\label{fig:bloch_sphere}
\end{figure}

In Fig.~\ref{fig:bloch_sphere} we plot the energy and energy variance densities on the Bloch sphere, as well as a complementary plot to Fig.~\ref{fig:v_epsilon} to show the density of Bloch states on the $(\En, v)$ surface.

In Table~\ref{tab:Bloch_sphere_variance} we have listed the energy and variance of the Bloch states we consider, as well as the thermal variance and entropy at the inverse temperature $\beta$ which results in a thermal energy equal to the state's energy. The states are mostly organized in equal energy series and sorted by decreasing variance. The thermal values were obtained from full spectrum data of $L=20$ sites.

\clearpage
\bibliography{bibliography}

\renewcommand{\thesection}{SM\arabic{section}}
\renewcommand{\thetable}{SM\arabic{table}}
\renewcommand{\thefigure}{SM\arabic{figure}}

\renewcommand{\Prefix}{SuppFigures} %
\clearpage
\onecolumngrid

\section*{Supplemental material}%
\label{sec:supplemental_material}

\begin{table}[h]
	\centering
	\begin{tabular}{CCCCCCCC}
		\Psi & \theta/\pi & \phi/\pi & \En     & v      & \tilde{v} & \beta   \\
		\toprule
		X_+  & 0.5        & 0        & -1.0500 & 1.2500 & 0.6130    & +0.7186 \\
		Z_+  & 0          & 0        & +1.5000 & 1.1025 & 0.9670    & -0.7275 \\
		E_-  & 0.552172   & 0        & -1.0909 & 0.9475 & 0.5080    & +0.7918 \\
		E_+  & -0.135159  & 0        & +1.7184 & 0.0288 & 0.0201    & -1.5975 \\
		\midrule
		Y_+  & 0.5        & 0.5      & 0       & 2.3525 & 2.3525    & -0.0000 \\
		y_1  & 0.27       & 0.071211 &         & 4.5498 &           &         \\
		y_2  & 0.335      & 0.318136 &         & 3.8344 &           &         \\
		y_3  & 0.4        & 0.419455 &         & 3.1615 &           &         \\
		y_4  & 0.465      & 0.479579 &         & 2.5961 &           &         \\
		y_6  & 0.53       & 0.511634 &         & 2.1820 &           &         \\
		y_7  & 0.595      & 0.51922  &         & 1.9373 &           &         \\
		\midrule
		z_1  & 0.2046     & 0.17081  &         & 4.3441 &           &         \\
		z_2  & 0.2846     & 0.419324 &         & 3.7471 &           &         \\
		z_3  & 0.3646     & 0.541178 &         & 3.0981 &           &         \\
		z_4  & 0.4446     & 0.620907 & +0.5000 & 2.4788 & 2.4795    & -0.2044 \\
		z_5  & 0.5246     & 0.66991  &         & 1.9540 &           &         \\
		z_6  & 0.6046     & 0.689459 &         & 1.5612 &           &         \\
		z_7  & 0.6846     & 0.681326 &         & 1.3068 &           &         \\
		Z_-  & 1          & 0        &         & 1.1025 &           &         \\
		\midrule
		a_1  & 0.23954    & 0        &         & 4.6537 &           &         \\
		a_2  & 0.3218     & 0.364508 &         & 3.8445 &           &         \\
		a_3  & 0.4018     & 0.479667 &         & 3.0785 &           &         \\
		a_4  & 0.4818     & 0.545186 & +0.1801 & 2.4397 & 2.4409    & -0.0750 \\
		a_5  & 0.5618     & 0.574644 &         & 1.9952 &           &         \\
		a_6  & 0.6418     & 0.571113 &         & 1.7627 &           &         \\
		b_1  & 0.223      & 0.135305 &         & 4.5187 &           &         \\
		b_2  & 0.303      & 0.39516  &         & 3.8278 &           &         \\
		b_3  & 0.383      & 0.513387 &         & 3.1166 &           &         \\
		b_4  & 0.463      & 0.586065 & +0.3500 & 2.4794 & 2.4794    & -0.1440 \\
		b_5  & 0.543      & 0.625346 &         & 1.9842 &           &         \\
		b_6  & 0.623      & 0.633627 &         & 1.6609 &           &         \\
		b_7  & 0.703      & 0.612991 &         & 1.5001 &           &         \\
		\bottomrule
		\Psi & \theta/\pi & \phi/\pi & \En     & v      & \tilde{v} & \beta
	\end{tabular}%
	\hfill%
	\begin{tabular}{CCCCCCCC}
		\Psi & \theta/\pi & \phi/\pi & \En     & v      & \tilde{v} & \beta   \\
		\toprule
		c_1  & -0.18      & 0.804752 &         & 4.0489 &           &         \\
		c_2  & -0.26      & 0.556808 &         & 3.5706 &           &         \\
		c_3  & -0.34      & 0.429533 &         & 3.0112 &           &         \\
		c_4  & -0.42      & 0.340407 & +0.6750 & 2.4345 & 2.4364    & -0.2755 \\
		c_5  & -0.5       & 0.277749 &         & 1.8969 &           &         \\
		c_6  & -0.58      & 0.241765 &         & 1.4397 &           &         \\
		c_7  & -0.66      & 0.23333  &         & 1.0842 &           &         \\
		c_8  & -0.74      & 0.245595 &         & 0.8324 &           &         \\
		c_9  & -0.82      & 0.26067  &         & 0.6716 &           &         \\
		\midrule
		d_1  & -0.151     & 0.795424 &         & 3.6393 &           &         \\
		d_2  & -0.231     & 0.537251 &         & 3.2877 &           &         \\
		d_3  & -0.311     & 0.403645 &         & 2.8378 &           &         \\
		d_4  & -0.391     & 0.303874 & +0.8522 & 2.3311 & 2.3325    & -0.3496 \\
		d_5  & -0.471     & 0.223441 &         & 1.8107 &           &         \\
		d_6  & -0.551     & 0.161157 &         & 1.3147 &           &         \\
		d_7  & -0.631     & 0.122279 &         & 0.8735 &           &         \\
		d_8  & -0.711     & 0.107143 &         & 0.5074 &           &         \\
		d_9  & -0.821446  & 0        &         & 0.1447 &           &         \\
		\midrule
		x_1  & 0.1295     & 0.291985 &         & 3.0024 &           &         \\
		x_2  & 0.2045     & 0.504394 &         & 2.7872 &           &         \\
		x_3  & 0.2795     & 0.630833 &         & 2.4865 &           &         \\
		x_4  & 0.3545     & 0.735193 & +1.0500 & 2.1132 & 2.1145    & -0.4382 \\
		x_5  & 0.4295     & 0.836335 &         & 1.6845 &           &         \\
		X_-  & 0.5        & 1        &         & 1.2500 &           &         \\
		\midrule
		l_1  & 0.7996     & 0.2692   & -0.1608 & 2.2269 & 2.2270    & +0.0701 \\
		l_2  & 0.7257     & 0.3281   & -0.3114 & 2.0652 & 2.0653    & +0.1402 \\
		l_3  & 0.6785     & 0.335    & -0.4236 & 1.9147 & 1.9147    & +0.1966 \\
		l_4  & 0.61       & 0.3391   & -0.5330 & 1.7427 & 1.7427    & +0.2564 \\
		l_5  & 0.608      & 0.29     & -0.6625 & 1.6163 & 1.5064    & +0.3361 \\
		l_6  & 0.607      & 0.24     & -0.7787 & 1.4735 & 1.2653    & +0.4200 \\
		l_7  & 0.6035     & 0.185    & -0.8893 & 1.3084 & 1.0124    & +0.5174 \\
		l_8  & 0.6        & 0.12     & -0.9875 & 1.1389 & 0.7718    & +0.6280 \\
		\bottomrule
		\Psi & \theta/\pi & \phi/\pi & \En     & v      & \tilde{v} & \beta
	\end{tabular}
	\caption{List of all initial Bloch states, including principal axes states ($X_{\pm}, Y_{\pm}, Z_{\pm}$), extremes of energy ($E_{\pm}$), local minimum ($d_9$) and global maximum ($a_1$) of the energy variance.}%
	\label{tab:Bloch_sphere_variance}
\end{table}

\begin{table}[h]
	\centering
	\begin{tabular}{CCCC}
		\Psi & l & \bar{I}_{l,l} (L=24)   & \tilde{I}_{l,l} (\tilde{L}=14) \\
		Y_+  & 1 & 9.13312 \times 10^{-7} & 2.48177 \times 10^{-8}         \\
		a_4  & 1 & 4.03883 \times 10^{-3} & 4.01908 \times 10^{-3}         \\
		z_4  & 1 & 2.78593 \times 10^{-2} & 2.76228 \times 10^{-2}         \\
		Y_+  & 2 & 1.65851 \times 10^{-5} & 3.51203 \times 10^{-8}         \\
		a_4  & 2 & 4.04465 \times 10^{-3} & 4.01919 \times 10^{-3}         \\
		z_4  & 2 & 2.79354 \times 10^{-2} & 2.76329 \times 10^{-2}         \\
	\end{tabular}%
	\hspace{1cm}%
	\begin{tabular}{CCCC}
		\Psi & l & \bar{I}_{l,l} (L=30)   & \tilde{I}_{l,l} (\tilde{L}=14) \\
		Y_+  & 1 & 4.24647 \times 10^{-8} & 0.65216 \times 10^{-9}         \\
		a_4  & 1 & 4.01980 \times 10^{-3} & 4.01904 \times 10^{-3}         \\
		b_4  & 1 & 1.43633 \times 10^{-2} & 1.43534 \times 10^{-2}         \\
		z_4  & 1 & 2.76776 \times 10^{-2} & 2.76228 \times 10^{-2}         \\
		c_4  & 1 & 4.64905 \times 10^{-2} & 4.62124 \times 10^{-2}         \\
		i_4  & 1 & 6.69230 \times 10^{-2} & 6.57969 \times 10^{-2}         \\
		x_4  & 1 & 8.88850 \times 10^{-2} & 8.42359 \times 10^{-2}         \\
	\end{tabular}
	\caption{List of mutual information values as plotted in Fig.~\ref{fig:I_E}.}%
	\label{tab:I_l_l}
\end{table}

\begin{figure}
	\centering
	\foreach \state in {y, a, b, z, c, d, x}
		{%
			\plot{O-Lv,L,psi=\state_i,O=C2.pdf}
		}
	\caption{Finite size deviations of $C_2$ as a function of the predicted ETH scaling $(v-\tilde{v})/L$ for several equal energy series of states. The dashed lines have slope equal to the thermal estimate $\DD \tilde{O}$ and cross the origin.}%
	\label{fig:Obar_Lv_C2}
\end{figure}

\begin{figure}
	\centering
	\foreach \state in {y, a, b, z, c, d, x}
		{%
			\plot{O-Lv,L,psi=\state_i,O=Sz0.pdf}
		}
	\caption{Finite size deviations of $\sigma^z$ as a function of the predicted ETH scaling $(v-\tilde{v})/L$ for several equal energy series of states. The dashed lines have slope equal to the thermal estimate $\DD \tilde{O}$ and cross the origin.}%
	\label{fig:Obar_Lv_Sz}
\end{figure}

\begin{figure}
	\centering
	\foreach \state in {y, a, b, z, c, d, x}
		{%
			\plot{O-Lv,L,psi=\state_i,O=SxSx0,1.pdf}
		}
	\caption{Finite size deviations of $\sigma^x_0\sigma^x_1$ as a function of the predicted ETH scaling $(v-\tilde{v})/L$ for several equal energy series of states. The dashed lines have slope equal to the thermal estimate $\DD \tilde{O}$ and cross the origin.}%
	\label{fig:Obar_Lv_SxSx}
\end{figure}

\begin{figure}
	\centering
	\foreach \state in {y, a, b, z, c, d, x}
		{%
			\plot{O-Lv,L,psi=\state_i,O=SzSz0,1.pdf}
		}
	\caption{Finite size deviations of $\sigma^z_0\sigma^z_1$ as a function of the predicted ETH scaling $(v-\tilde{v})/L$ for several equal energy series of states. The dashed lines have slope equal to the thermal estimate $\DD \tilde{O}$ and cross the origin.}%
	\label{fig:Obar_Lv_SzSz}
\end{figure}

\begin{figure}
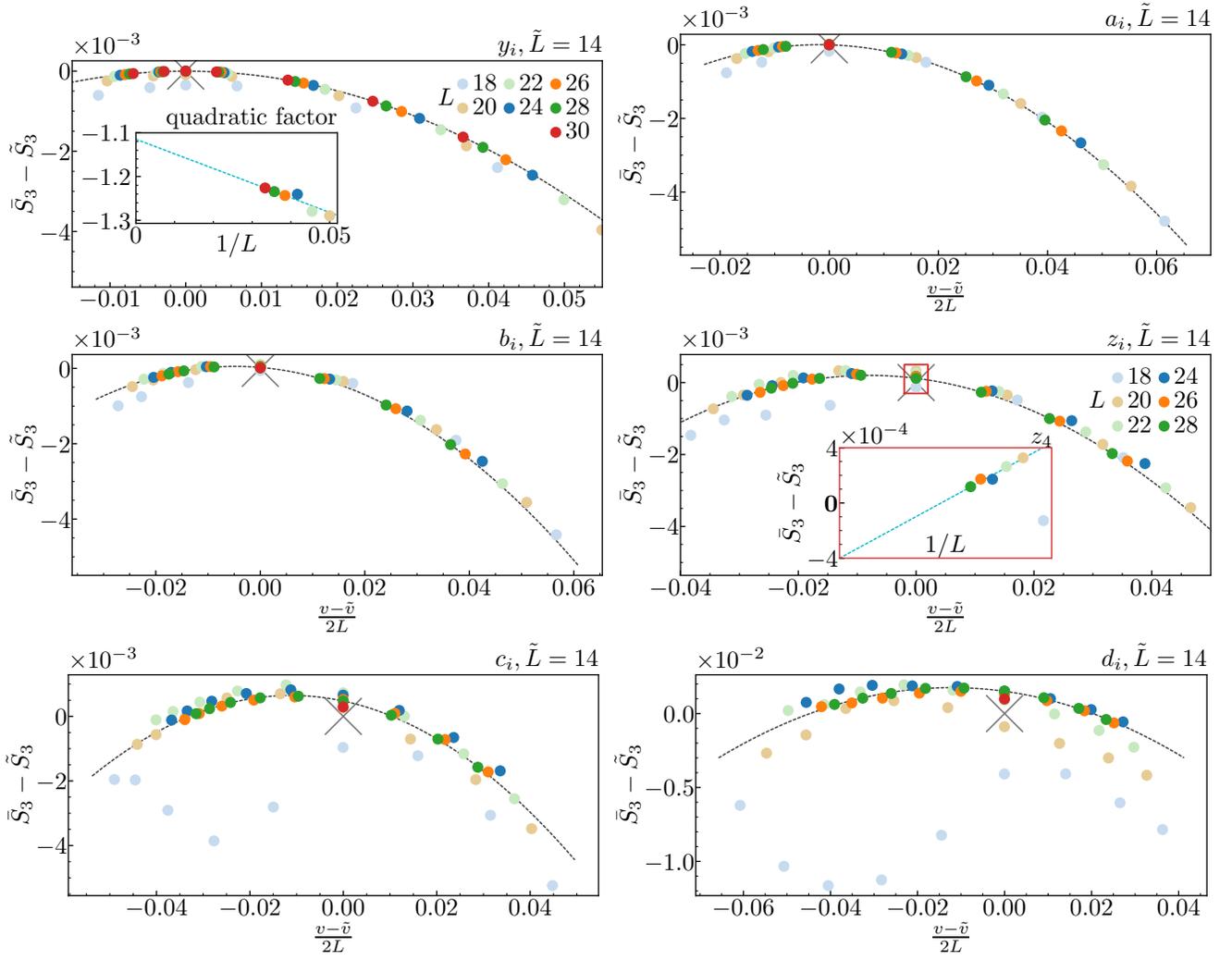

	\centering
	\foreach \state in {y, a, b, z, c, d}
		{%
			\plot{O-Lv,L,psi=\state_i,O=S0,1,2.pdf}
		}
	\caption{Finite size deviations between the entanglement entropy of a three-site cluster and the rescaled thermal entropy, for several equal energy series of states. The origin is marked with a $\times$.}%
	\label{fig:S_Lv_sup}
\end{figure}

\begin{figure}
	\centering
	\foreach \state in {
			y_1, y_2, y_3, y_4, Y_+, y_6, y_7,
			a_1, a_2, a_3, a_4, a_5, a_6}
		{%
			\plot{logO-t,L,psi=\state,O=C2.pdf}
		}
	\caption{Exponential decay of the energy correlator $C_2$ on a linear-log scale, for the $y_i$, and $a_i$ series of states, with fitted exponentials shown with black dashed lines. Two black dots along the fitted lines mark the limits of the time interval where the fitting was performed.}%
	\label{fig:logO_t_1}
\end{figure}

\begin{figure}
	\centering
	\foreach \state in {
			y_1, y_2, y_3, y_4, Y_+, y_6, y_7,
			a_1, a_2, a_3, a_4, a_5, a_6}
		{%
			\plot{logO-t,L,shifted,psi=\state,O=C2.pdf}
		}
	\caption{Exponential decay of the energy correlator $C_2$ on a linear-log scale, for the $y_i$, and $a_i$ series of states, with the fitted exponentials shown with black dashed lines. These are the same data lines shown in Fig.~\ref{fig:logO_t_1}, except they have been shifted from each other for better visualization as they have a significant overlap ($L$ is increasing from bottom to top). Two black dots along the fitted lines mark the limits of the time interval where the fitting was performed.}%
	\label{fig:logO_t_shifted_1}
\end{figure}

\begin{figure}
	\centering
	\foreach \state in {
			b_1, b_2, b_3, b_4, b_5, b_6,
			z_1, z_2, z_3, z_4, z_5, z_6, z_7, Z_-}
		{%
			\plot{logO-t,L,psi=\state,O=C2.pdf}
		}
	\caption{Exponential decay of the energy correlator $C_2$ on a linear-log scale, for the $b_i$, and $z_i$ series of states, with fitted exponentials shown with black dashed lines. Two black dots along the fitted lines mark the limits of the time interval where the fitting was performed.}%
	\label{fig:logO_t_2}
\end{figure}

\begin{figure}
	\centering
	\foreach \state in {
			b_1, b_2, b_3, b_4, b_5, b_6,
			z_1, z_2, z_3, z_4, z_5, z_6, z_7, Z_-}
		{%
			\plot{logO-t,L,shifted,psi=\state,O=C2.pdf}
		}
	\caption{Exponential decay of the energy correlator $C_2$ on a linear-log scale, for the $b_i$ and $z_i$ series of states, with the fitted exponentials shown with black dashed lines. These are the same data lines shown in Fig.~\ref{fig:logO_t_2}, except they have been shifted from each other for better visualization as they have a significant overlap ($L$ is increasing from bottom to top). Two black dots along the fitted lines mark the limits of the time interval where the fitting was performed.}%
	\label{fig:logO_t_shifted_2}
\end{figure}

\begin{figure}
	\centering
	\foreach \state in {
			c_1, c_2, c_3, c_4, c_5, c_6, c_7, c_8, c_9,
			d_1, d_2, d_3, d_4, d_5, d_6, d_7, d_8, d_9}
		{%
			\plot{logO-t,L,psi=\state,O=C2.pdf}
		}
	\caption{Exponential decay of the energy correlator $C_2$ on a linear-log scale, for the $c_i$, and $d_i$ series of states, with fitted exponentials shown with black dashed lines. Two black dots along the fitted lines mark the limits of the time interval where the fitting was performed.}%
	\label{fig:logO_t_3}
\end{figure}

\begin{figure}
	\centering
	\foreach \state in {
			c_1, c_2, c_3, c_4, c_5, c_6, c_7, c_8, c_9,
			d_1, d_2, d_3, d_4, d_5, d_6, d_7, d_8, d_9}
		{%
			\plot{logO-t,L,shifted,psi=\state,O=C2.pdf}
		}
	\caption{Exponential decay of the energy correlator $C_2$ on a linear-log scale, for the $c_i$ and $d_i$ series of states, with the fitted exponentials shown with black dashed lines. These are the same data lines shown in Fig.~\ref{fig:logO_t_3}, except they have been shifted from each other for better visualization as they have a significant overlap ($L$ is increasing from bottom to top). Two black dots along the fitted lines mark the limits of the time interval where the fitting was performed.}%
	\label{fig:logO_t_shifted_3}
\end{figure}

\end{document}